\documentclass[journal]{IEEEtran}

\usepackage{fixltx2e}
\usepackage{amssymb}
\usepackage[cmex10]{amsmath}
\usepackage{amsfonts}
\usepackage{lineno}
\modulolinenumbers[5]
\usepackage{url}
\usepackage{hyperref}
\usepackage[plain]{fancyref}
\usepackage{textgreek}
\usepackage{tikz}
\usepackage{graphicx}
\usepackage{grffile}
\graphicspath{{./fig/}}
\DeclareGraphicsExtensions{.pdf,.jpeg,.jpg,.png}
\usepackage{xcolor}
\usepackage{array}
\usepackage{multirow}
\usepackage{tabulary}
\usepackage{booktabs}
\usepackage{setspace}
\newcommand{\isot}[2]{\(^{#2}\){#1}}

\begin{document}
\IEEEpubid{\begin{minipage}{0.85\textwidth}\ \\[12pt]\\ \\ \\ \centering
  \textcopyright 2020 IEEE.  Personal use of this material is permitted. Permission from IEEE must be obtained for all other uses, in any current or future media, including reprinting/republishing this material for advertising or promotional purposes, creating new collective works, for resale or redistribution to servers or lists, or reuse of any copyrighted component of this work in other works.
\end{minipage}}

\title{Modeling Aerial Gamma-Ray Backgrounds using Non-negative Matrix Factorization}

\author{M.~S.~Bandstra,
        T.~H.~Y.~Joshi,
        K.~J.~Bilton,
        A.~Zoglauer,
        and B.~J.~Quiter%
\thanks{M.~S.~Bandstra, T.~H.~Y.~Joshi, and B.~J.~Quiter are with the Nuclear Science Division at Lawrence Berkeley National Laboratory, Berkeley, CA 94720 USA e-mail: msbandstra@lbl.gov.}%
\thanks{K.~J.~Bilton is with the Department of Nuclear Engineering at the University of California, Berkeley, Berkeley, CA 94720 USA.}%
\thanks{A.~Zoglauer is with the Berkeley Institute for Data Science and the Space Sciences Laboratory at the University of California, Berkeley, Berkeley, CA 94720 USA.}%
\thanks{Manuscript received ---\@. Revised ---.}}

\markboth{IEEE Transactions on Nuclear Science}%
{---}

\maketitle

\begin{abstract}
Airborne gamma-ray surveys are useful for many applications, ranging from geology and mining to public health and nuclear security.
In all these contexts, the ability to decompose a measured spectrum into a linear combination of background source terms can provide useful insights into the data and lead to improvements over techniques that use spectral energy windows.
Multiple methods for the linear decomposition of spectra exist but are subject to various drawbacks, such as allowing negative photon fluxes or requiring detailed Monte Carlo modeling.
We propose using Non-negative Matrix Factorization (NMF) as a data-driven approach to spectral decomposition.
Using aerial surveys that include flights over water, we demonstrate that the mathematical approach of NMF finds physically relevant structure in aerial gamma-ray background, namely that measured spectra can be expressed as the sum of nearby terrestrial emission, distant terrestrial emission, and radon and cosmic emission.
These NMF background components are compared to the background components obtained using Noise-Adjusted Singular Value Decomposition (NASVD), which contain negative photon fluxes and thus do not represent emission spectra in as straightforward a way.
Finally, we comment on potential areas of research that are enabled by NMF decompositions, such as new approaches to spectral anomaly detection and data fusion.
\end{abstract}

\IEEEpeerreviewmaketitle%



\section{Introduction}
\IEEEPARstart{A}{erial} surveys are used to measure gamma-ray emission over large areas, with applications in geological exploration, radiation protection, mapping distributed contamination, and finding radiological sources outside of regulatory control~\cite{moxham_airborne_1960, galbraith_rock_1983, sanderson_aerial_1991, sanada_aerial_2015}.
Commonly used analysis techniques utilize the counts in certain spectral energy windows to measure and map background levels or to detect sources, and these approaches generally trade statistics for specificity to particular radioactive isotopes~\cite{grasty_uranium_1975, grasty_analysis_1985, detwiler_spectral_2015}.
Other techniques have been developed to leverage the information contained across the entire spectrum, thus gaining statistics but potentially losing specificity~\cite{crossley_inversion_1982, minty_airborne_1992, minty_multichannel_1998, aage_new_1999}.
Though potentially more powerful than windowed approaches, these techniques require detailed modeling of the different background components in order to compensate for the loss of specificity.
Here we will review current full-spectrum approaches and present a new approach using Non-negative Matrix Factorization (NMF) that, due to the reformulation of the problem, provides new insights into airborne survey measurements.


\subsection{Aerial gamma-ray background}
The natural backgrounds encountered by airborne systems are a combination of terrestrial sources and non-terrestrial sources (e.g.,~\cite{minty_fundamentals_1997}).
The terrestrial sources are typically a combination of \isot{K}{40}, \isot{U}{238}/\isot{U}{235} decay series isotopes, and \isot{Th}{232} decay series isotopes, collectively denoted KUT backgrounds.
Emission from terrestrial sources can be classified into two types: \textit{direct} emission, consisting of unscattered and scattered photons that are incident on the detector from below, and \textit{skyshine}, consisting of photons that have scattered in the air above the detector that are incident from above~\cite{minty_multichannel_1998}.
The non-terrestrial backgrounds are radon gas and its radioactive progeny, the effects of cosmic rays, and KUT backgrounds from the aircraft itself.
The radon component consists primarily of \isot{Rn}{222} progeny, and the cosmic component contains \(511\)~keV emission from cosmogenic positron production and a cosmogenic power-law continuum~\cite{sandness_accurate_2009, mitchell_skyshine_2009}.

An illustrative study of the full-spectrum background components for a stationary ground-based detector was undertaken in refs.~\cite{sandness_accurate_2009, mitchell_skyshine_2009}.
In that study the authors empirically separate terrestrial from non-terrestrial backgrounds, isolate skyshine from other terrestrial emission, and generate models that correctly reproduce each component.
Fully modeling the background required Monte Carlo simulations with terrestrial emission up to at least \(300\)~m away from the detector and with \(300\)~m of atmosphere overhead to produce the correct amount of skyshine, showing the complexity of the problem for even a static detector on the ground.
Other ground-based studies have shown the influence of skyshine at low energies even for distant sources~\cite{beck_radiation_1968, nason_benchmark_1981}.


\subsection{Background models}
Before going any further, we note that the term \textit{spectrum model} will carry a dual meaning in this work.
In the first sense, a spectrum model is a Monte Carlo physics simulation that reproduces the components of a measured spectrum.
In the second sense, a spectrum model is any mathematical decomposition of a measured spectrum, regardless of physical meaning of the components.
Both models provide a numerical description of the data, with the former providing a physical understanding and the latter providing a mathematical understanding.
An ideal method would include aspects of both kinds of models in one physical and mathematical description.

For a mathematical model of gamma-ray background to be consistent with physics (e.g., the additive nature of photons), the measured spectra should be expressed as a linear sum of spectral components.
For this work, we begin with a measured dataset \(\mathbf{X}\), which is an \(n \times d\) matrix where each of the \(n\)~rows is a measured spectrum with \(d\)~bins.
We consider linear decompositions of \(\mathbf{X}\) of the form
\begin{equation}\label{eq:linear_decomp}
    \mathbf{X} \approx \mathbf{\hat{X}} = \mathbf{A} \mathbf{V},
\end{equation}
where each row of the \(n \times d\) matrix \(\mathbf{\hat{X}}\) is a model-estimated spectrum, \(\mathbf{A}\) is an \(n \times k\) matrix of weights, \(\mathbf{V}\) is a \(k \times d\) matrix whose rows are the component spectra, and \(k \leq d\) is the number of components.
In nearly all cases, \fref{eq:linear_decomp} is an ill-posed problem with many potential solutions that depend on the choice of \(k\) and desired constraints on \(\mathbf{A}\) and \(\mathbf{V}\).
Here we will review some of the approaches for finding solutions, along with their advantages and disadvantages.


\subsection{Full Spectrum Analysis}
Full Spectrum Analysis (FSA) of gamma-ray spectra is one approach that has been studied for aerial and borehole detector systems~\cite{crossley_inversion_1982, minty_airborne_1992, minty_multichannel_1998, smith_multi-function_1983, hendriks_full-spectrum_2001}.
FSA aims to express a measured spectrum as a linear combination of component spectra that are predetermined through simulations and calibration measurements.
For example, component spectra for the KUT backgrounds have been developed to  determine the isotopic compositions of the ground beneath an aircraft~\cite{hendriks_full-spectrum_2001}.
The \(k\) spectral components make up the rows of \(\mathbf{V}\), and the weights \(\mathbf{A}\) are typically calculated by minimizing the sum of the square differences between \(\mathbf{X}\) and \(\mathbf{\hat{X}}\), but since least squares can admit non-physical (i.e., negative) solutions to~\fref{eq:linear_decomp}, using non-negative least squares has been suggested to enforce the non-negativity of \(\mathbf{A}\)~\cite{caciolli_new_2012}.
Another desirable property for a spectrum model is to preserve the number of counts in each spectrum, which least squares minimization alone does not enforce, but this constraint can be included in the minimization~\cite{caciolli_new_2012}.

One challenge of FSA is that the decomposition can be poor if the component spectra are not correct, so care must be taken to ensure their accuracy.
The component spectra are usually determined by using a combination of Monte Carlo simulations and calibration measurements, for example by measuring slabs of material with known KUT concentrations underneath wooden boards to simulate direct terrestrial emission (e.g.,~\cite{dickson_utilizing_1981}).
However, such measurements are unable to reproduce the effects of skyshine, which is a major terrestrial contribution to airborne spectra below \(400\)~keV~\cite{minty_multichannel_1998}.


\subsection{Noise-Adjusted Singular Value Decomposition}
Other full-spectrum modeling techniques have focused on using mathematical methods to discover and utilize structure found within measured data without requiring a physical understanding.
One commonly used method is Noise-Adjusted Singular Value Decomposition (NASVD), which was developed to remove statistical noise from aerial gamma-ray measurements by attempting to separate true spectral variations from statistical noise~\cite{hovgaard_new_1997, hovgaard_reducing_1997, minty_improved_1998}.
NASVD improved upon similar work using Principal Component Analysis (PCA)~\cite{dickson_utilizing_1981}.
NASVD has been used to map low levels of contamination~\cite{aage_new_1999} and has recently been applied to the problem of estimating real-time backgrounds for aerial surveys~\cite{kulisek_real-time_2015}.
NASVD decomposes measured spectra into principal components and then keeps only those components which represent true spectral variability.
Although NASVD is typically used to smooth spectra before further processing, isotope-specific signatures in the first few components have led some to comment on the possible associations between NASVD components and KUT and radon backgrounds (e.g.,~\cite{hovgaard_reducing_1997}).
However, unlike FSA, all but the first component contain negative values, so this method permits solutions to~\fref{eq:linear_decomp} that have negative photon fluxes, which are non-physical.


\subsection{Non-negative Matrix Factorization}
Recently Non-negative Matrix Factorization (NMF)~\cite{paatero_positive_1994, lee_learning_1999} was introduced as a technique for performing dimensionality reduction of gamma-ray background data~\cite{bilton_non-negative_2019} and for gamma-ray source identification~\cite{koudelka_modular_2016}.
NMF has already been used for spectral deconvolution in other fields of spectroscopy (e.g.,~\cite{sajda_nonnegative_2004, liou_unsupervised_2005, pauca_nonnegative_2006}).
NMF is a method of solving~\fref{eq:linear_decomp} such that both \(\mathbf{A}\) and \(\mathbf{V}\) are constrained to be non-negative.
NMF can be thought of as a bridge between FSA and NASVD --- like NASVD, the NMF components are determined from the measurements themselves, and like FSA, the NMF components are compatible with physics and thus are capable (although not guaranteed) of describing actual background emissions.
Indeed recent analysis of data from a vehicle-borne system has found correlations between NMF-derived background components and features derived from panoramic images of the scene around the vehicle~\cite{bandstra_attribution_2018}.
Other work has shown NMF to have advantages over other methods for spectral anomaly detection and identification~\cite{bilton_non-negative_2019}.

In this work NMF is applied to aerial survey data and the resulting components are interpreted as known background sources, thus connecting a purely mathematical model of the spectra with a physical model.
We start in \Fref{sec:lakemohave} by analyzing a survey containing passes over a land-water interface at low altitude to demonstrate the basic method.
NMF components are compared to a full Monte Carlo model in \Fref{sec:lakemohave_modeling} and \Fref{sec:lakemohave_model_results}, and then compared with NASVD components in \Fref{sec:nasvd}.
More complex datasets are analyzed in \Fref{sec:bayarea}, showing the potentially wide applicability of the method.
Finally, the implications for mapping, anomaly detection, and data fusion using NMF are discussed in \Fref{sec:discuss}.


\section{Analysis of a land-water interface}\label{sec:lakemohave}
\begin{figure}[t!]
    \centering
    \begin{tikzpicture}
        \node[anchor=south west,inner sep=0] (image) at (0,0) {\includegraphics[width=\columnwidth]{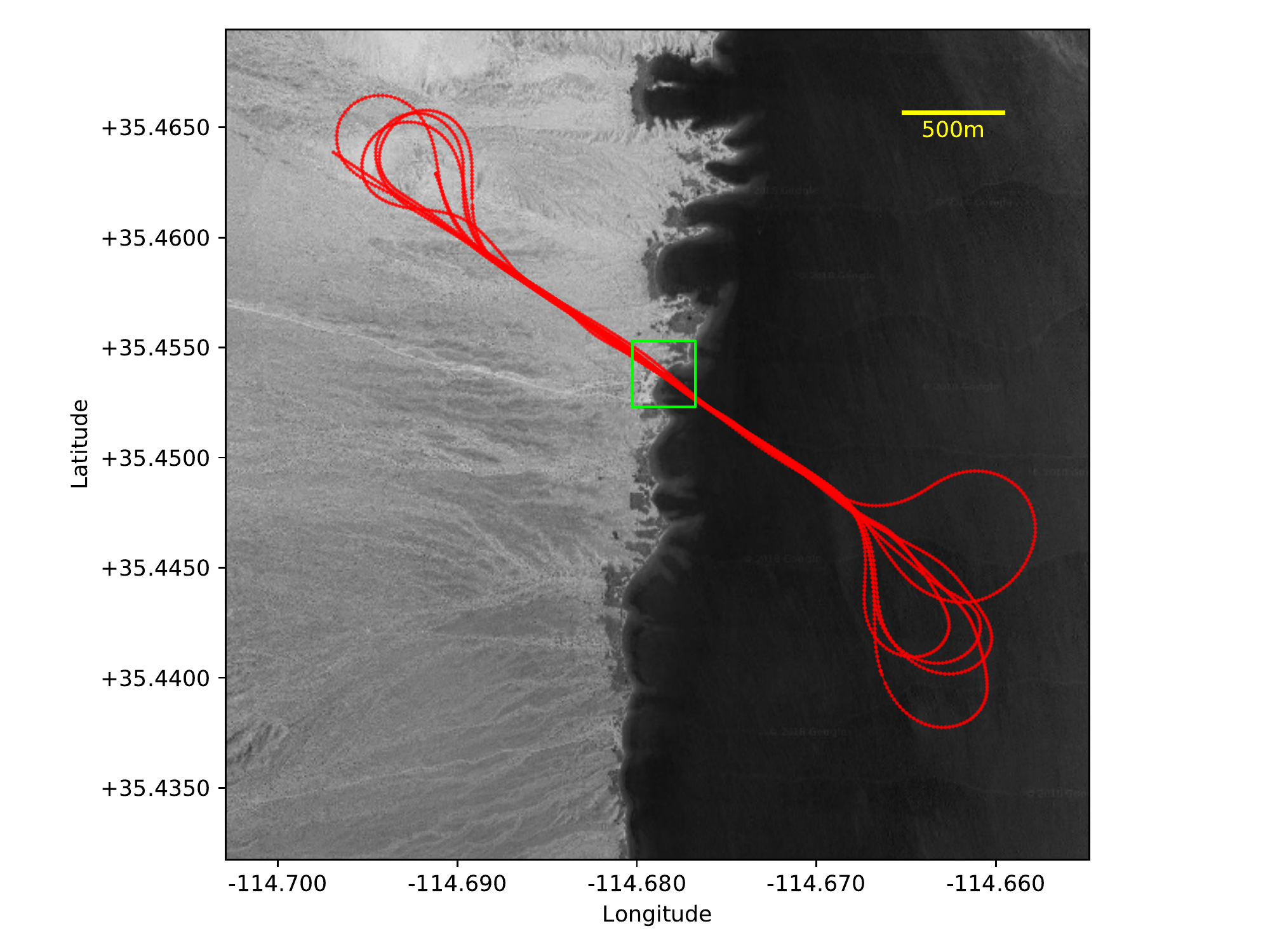}};
        \begin{scope}[x={(image.south east)},y={(image.north west)}]
            \node[anchor=south west,inner sep=0] (image) at (0.19,0.13) {\fcolorbox{green}{green}{\includegraphics[width=0.25\columnwidth]{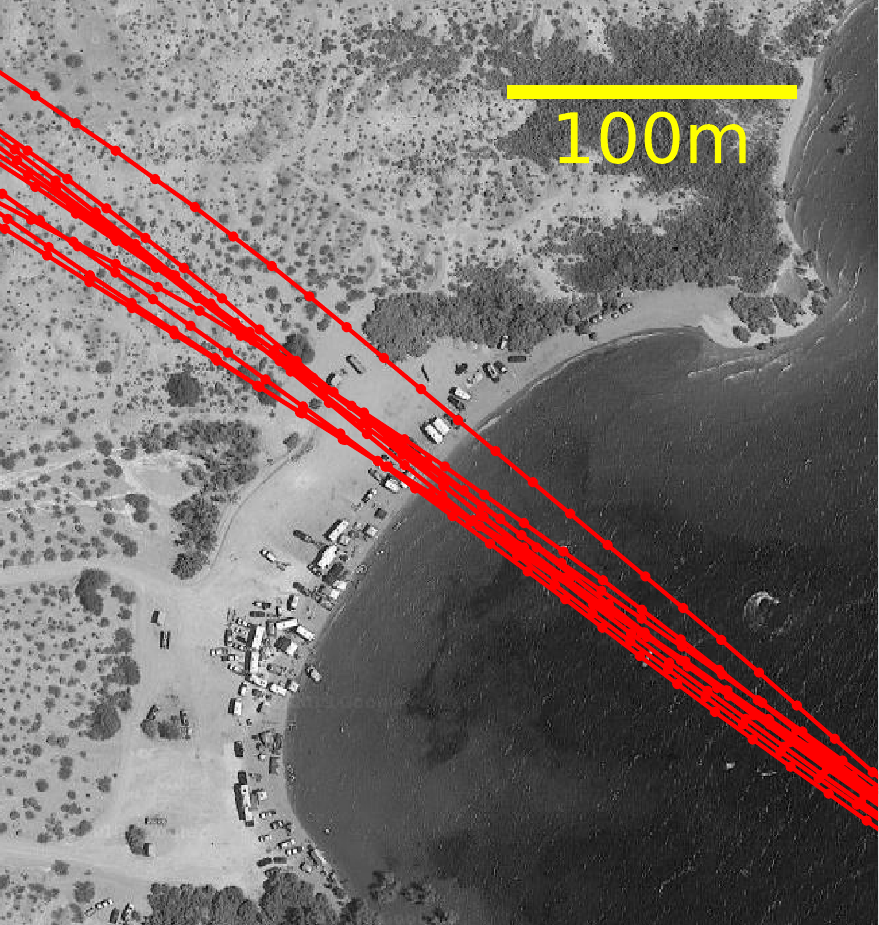}}};
        \end{scope}
    \end{tikzpicture}
\caption{The helicopter flight path of the Lake Mohave dataset showing repeated land-water crossings.
The inset shows a magnified view of the shoreline where the land-water crossings were made.
(Map imagery: Google.)\label{fig:lakemohave_map}}
\end{figure}

The data analyzed here were taken on 14~February~2012 at Lake Mohave on the Nevada-Arizona border.
The detector system was flown on a helicopter and consisted of four Radiation Solutions Inc.~RSX-3 units, totaling twelve \(2 \times 4 \times 16\)-inch NaI(Tl) detectors~\cite{radiation_solutions_inc._airborne_2019}.
Photon events were recorded by the data acquisition system in \(3\)~keV bins from~\(0\) to \(3072\)~keV, and for this analysis an energy cut of \(30-3060\)~keV was used.
Events were rebinned into \(d = 200\)~non-uniform bins with bin widths approximately proportional to the square root of energy in order to approximate the detector energy resolution, decrease the sparsity of the spectral counts, and reduce the computation time needed for NMF\@.
The time range of interest is 12:12:15 to 12:32:05 local time for a total of \(n = 2381\)~samples at a rate of \(2\)~Hz.
During the collection, the aircraft made ten passes over a land-water interface at a nominal altitude of \(100\)~ft above ground level (\Fref{fig:lakemohave_map}).

\fboxsep=0pt
\fboxrule=1pt

Because of the repeated passes over the land-water interface, this dataset isolates a special case of background variation often encountered in aerial surveys.
Over water and far from land, the system essentially measures only radon and cosmic backgrounds, plus any aircraft background, while over land it measures a combination of all background sources.
Using this dataset, we are able to investigate the major features of the background and match these features to possible source terms.


\subsection{NMF decomposition method}\label{sec:lakemohave_decomp}
NMF reduces the dimensionality of data by solving~\fref{eq:linear_decomp} subject to the constraint that all of the matrices are non-negative.
Essentially, NMF decomposes each spectrum into a linear combination of a pre-specified (and in our case, small) number of constituent spectra using non-negative coefficients.
In reality, the detection of photon events by a detector should be consistent with this formulation since photons themselves are non-negative and, excepting the effects of detector dead time, detected photon events sum together linearly.

Interpreting the estimate \(\mathbf{\hat{X}} = \mathbf{A} \mathbf{V}\) as the mean of the Poisson-distributed photon event counts \(\mathbf{X}\), the objective function to minimize is the Poisson negative log likelihood
\begin{equation}\label{eq:loss}
    \Lambda(\mathbf{X} | \mathbf{\hat{X}}) = \sum_i \sum_j \left(\hat{X}_{ij} - X_{ij} \log(\hat{X}_{ij}) + \log X_{ij}!\right).
\end{equation}
There is no closed form solution to this problem so an iterative approach must be taken.
We use the maximum likelihood update rules given in ref.~\cite{lee_learning_1999} to perform the optimization over both \(\mathbf{A}\) and \(\mathbf{V}\), which will be referred to as \textit{training} the NMF model.
During the training, the rows of \(\mathbf{V}\) are constrained to sum to unity so that the weights \(\mathbf{A}\) will be the integrated number of photon counting events attributed to each component.

Two caveats to NMF are that the decomposition may not be unique and the decomposition may only find a local minimum.
If there exists a \(k \times k\) invertible matrix \(\mathbf{D}\) such that \( \mathbf{V'} \equiv \mathbf{D} \mathbf{V}\) and \( \mathbf{A'} \equiv \mathbf{A} \mathbf{D}^{-1}\) are both non-negative, then \(\mathbf{A'} \mathbf{V'} = \mathbf{A} \mathbf{V}\) is an equivalent decomposition~\cite{paatero_positive_1994}.
(Note that \(\mathbf{D}\) itself need not be non-negative.)
A trivial solution for \(\mathbf{D}\) is a permutation matrix, which underscores the fact that there is no preferred ordering of NMF components, unlike in PCA or SVD\@.
The introduction of additional constraints on \(\mathbf{A}\) and \(\mathbf{V}\) in the form of regularization terms helps to break the symmetry and guide the solutions to have desired properties~\cite{pauca_nonnegative_2006, yu_minimum_2007}, though no constraints are added here.
Since the ordering of components is arbitrary, we chose to sort the components in decreasing order of the variance of their weights, so that component~0 has the highest variability, component~1 has less variability, etc.
This ordering is analogous to the ordering that results from PCA or NASVD\@.

The choice of initialization of \(\mathbf{A}\) and \(\mathbf{V}\) can strongly influence the particular solution that the minimization algorithm finds, since NMF is a non-convex optimization problem and many optimization techniques will only find the nearest local minimum~\cite{lee_algorithms_2001}.
Many random initializations of \(\mathbf{A}\) and \(\mathbf{V}\) were initially tried, and although some physical features (that will be discussed later) were apparent, these components were often not smooth and therefore unlikely to be comparable with the spectra of different background sources.

For the sake of repeatability, and also because this initialization has generally led to smooth spectral components, we performed the following initialization.
We made an ansatz that NMF might at least separate the over-water background from the terrestrial background, and to estimate the over-water background we chose all spectra with total counts less than an arbitrary threshold of~\(1.05\) times the minimum.
The average total counts of this set were used to initialize one column of \(\mathbf{A}\), and the average unit-normalized spectrum was used to initialize the corresponding row of \(\mathbf{V}\).
The residual total counts were split evenly among the remaining \(k - 1\) components, and the corresponding rows of \(\mathbf{V}\) were initialized to the average unit-normalized spectrum of the residual spectra.
This initialization means that the remaining \(k - 1\) components would be initialized identically, which would result in mathematically identical treatment during the optimization process.
To distinguish these \(k - 1\) components from each other, small positive random numbers less than~\(10^{-6}\) were added to the remaining \(k - 1\) components of \(\mathbf{V}\).

\begin{figure}[t!]
\centering
\includegraphics[width=0.9\columnwidth]{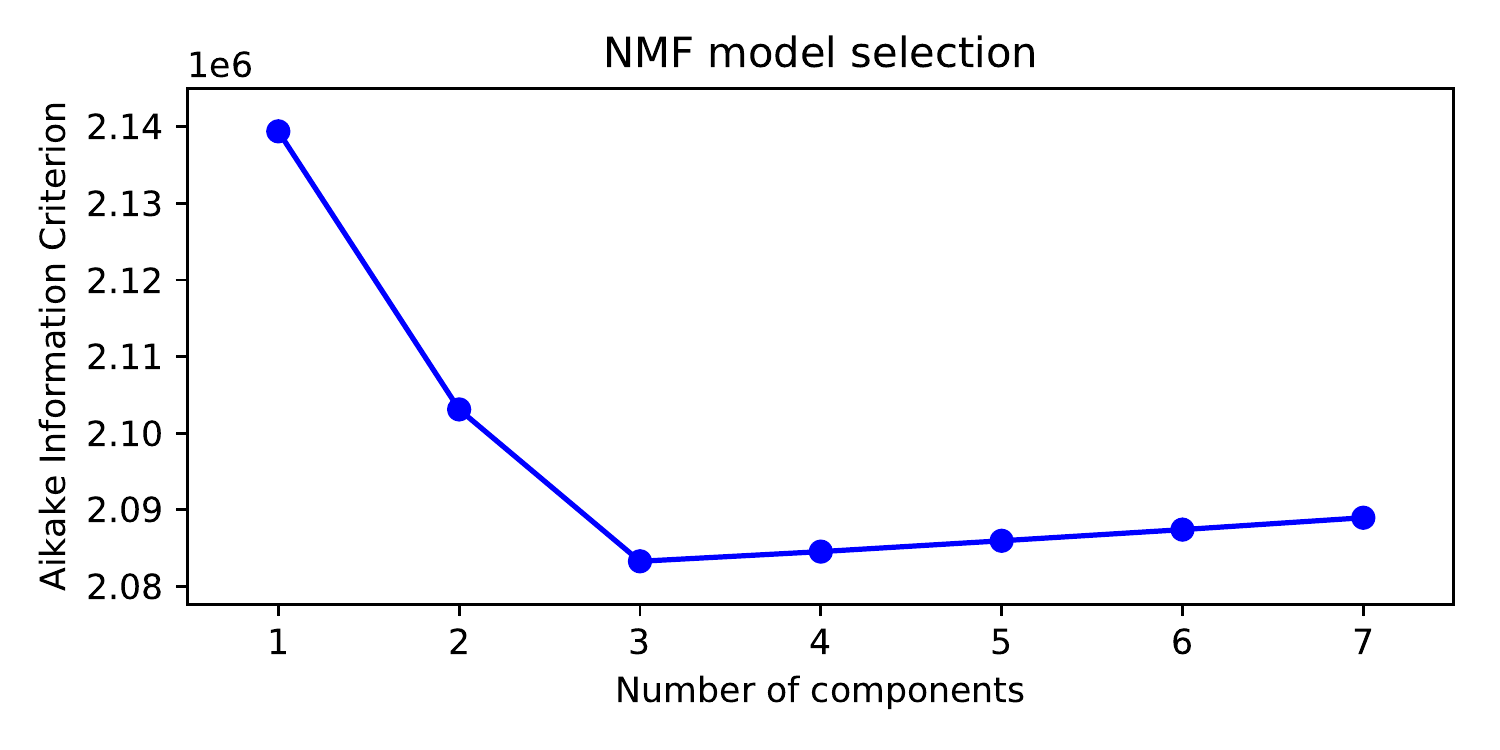}
\caption{Akaike Information Criterion (AIC) for NMF models with different numbers of components for the Lake Mohave dataset.
The model with three components has the best (lowest) AIC\@.
The line connecting the points has been added to guide the eye.\label{fig:aic}}
\end{figure}

As mentioned above, NMF differs from SVD in that the number of spectral components \(k\) must be chosen before calculating the decomposition.
NMF models with \(k=1\)~to~\(7\) components were fit to the Lake Mohave dataset, stopping the iterations after the change in \(\Lambda / n\) became less than an arbitrarily chosen threshold of~\(10^{-9}\).
To select the optimal number of components for a particular dataset, the Akaike Information Criterion (AIC)~\cite{akaike_new_1974} was used, as in related work~\cite{bilton_non-negative_2019}.
The AIC balances the likelihood of a model with the tendency to overfit as the number of model parameters is increased.
We find that for the dataset under examination here, three components is the optimal number according to AIC (\Fref{fig:aic}).

\subsection{NMF results}
\begin{figure*}[t!]
\centering
\includegraphics[width=0.42\textwidth]{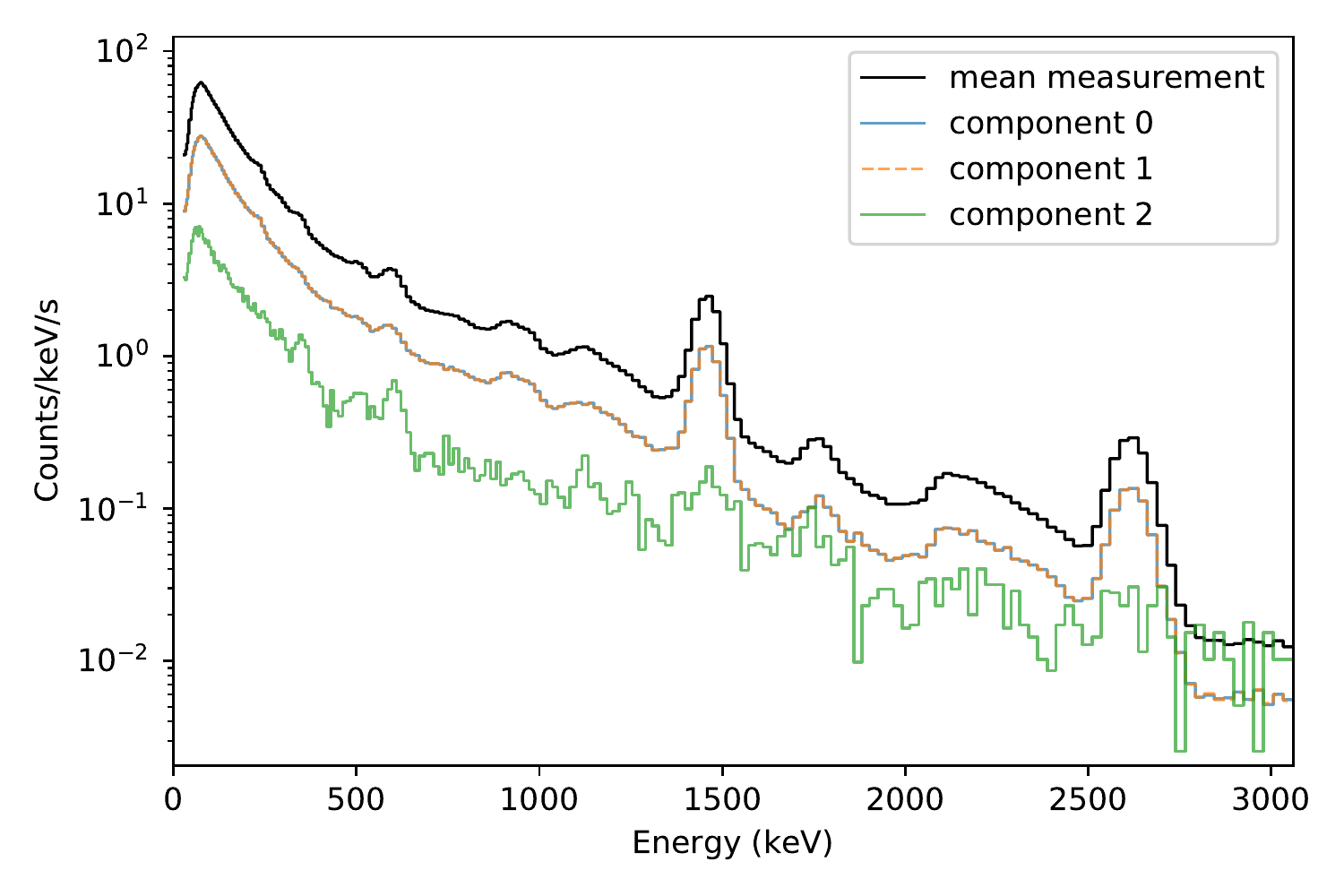}
\includegraphics[width=0.42\textwidth]{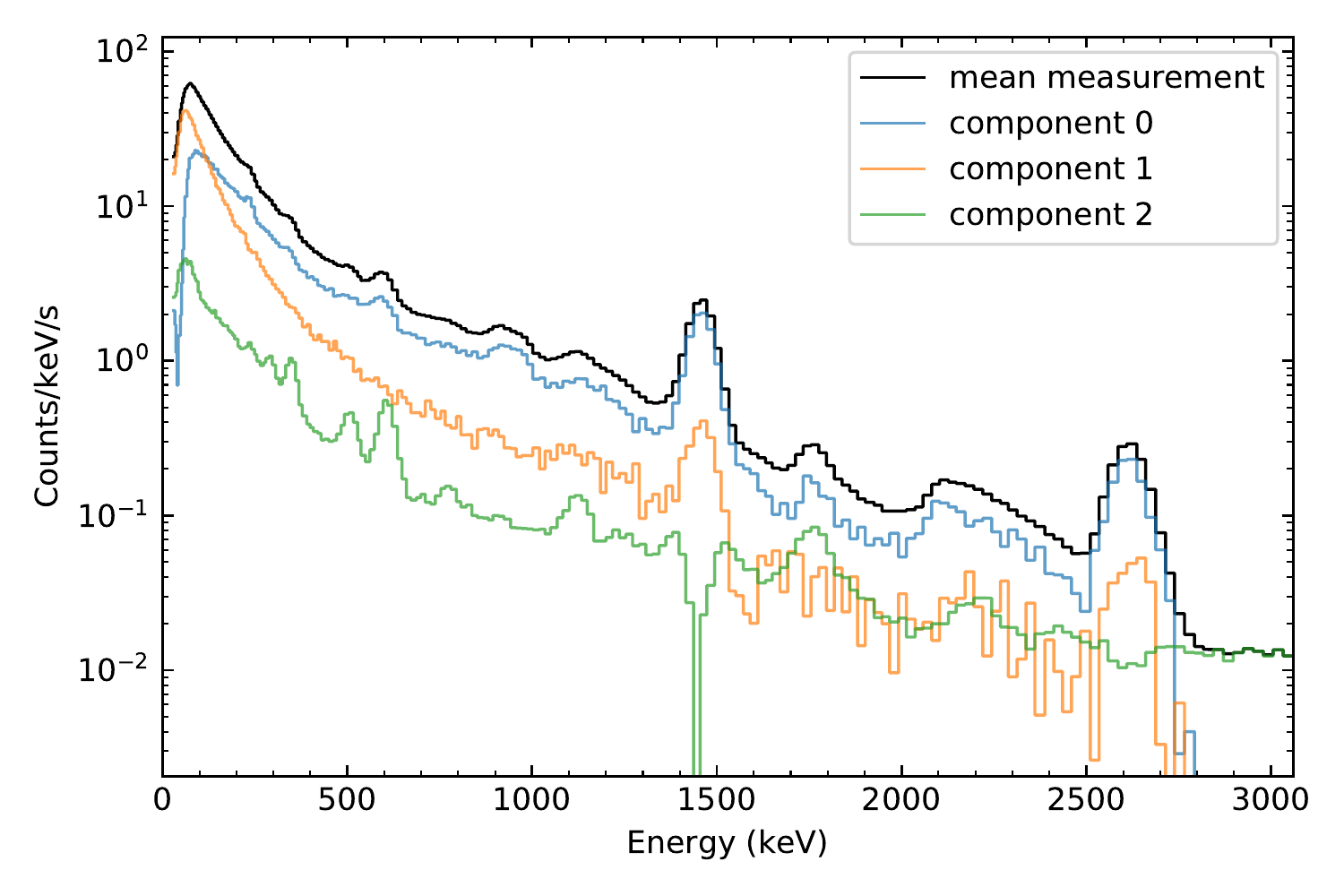}\\
\includegraphics[width=0.85\textwidth]{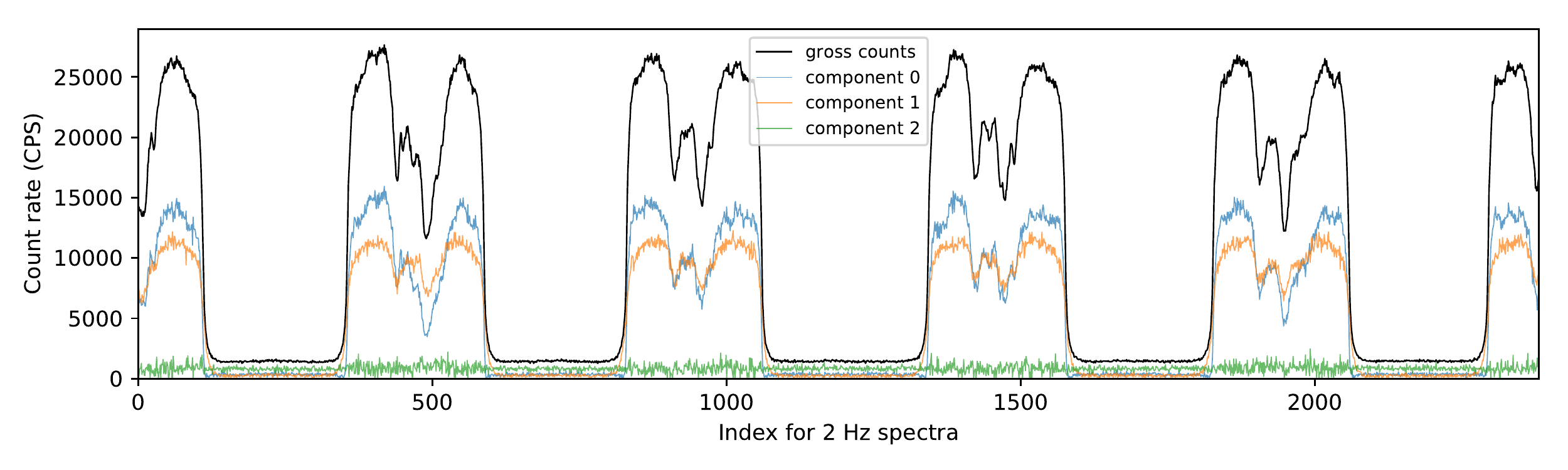}
\caption{NMF components and weights for a three-component decomposition of the Lake Mohave dataset.
The two top panels show the components initialized according to the procedure in the text (top left; components~0 and~1 are nearly identical) and the final components after optimization (top right).
The bottom panel shows the gross count rate and NMF component weights.
The components in the top plots are the rows of \(\mathbf{V}\) after dividing by bin widths and time, and then multiplying by their average weight to show their scale relative to the mean spectrum and each other.
The component weights are simply the columns of \(\mathbf{A}\).\label{fig:lakemohave_decomp}}
\end{figure*}

The initialized set of components for the three-component NMF decomposition and the resulting components after NMF training are shown in the upper pane of \Fref{fig:lakemohave_decomp}, along with the weights in the lower pane.
A comparison between the initialized components and the final components reveals that besides the engineered feature of an over-water component (component~2), the two other components have diverged significantly in shape from each other.
Some physical features are immediately apparent.

Component~0 is characterized by prominent \(1460.8\) and \(2614.5\)~keV lines from the decay of \isot{K}{40} and \isot{Tl}{208} (\isot{Th}{232} series), respectively, and several weaker lines are apparent: \(238.6\), \(338.3\), \(583.2\), \(911.2\), and \(969.0\)~keV, all of which are from the \isot{Th}{232} decay series.
This evidence points to nearby terrestrial emission as a possible source for component~0 events.

Component~1, on the other hand, is characterized by a high rise at the low energy end of the spectrum, peaking below \(100\)~keV, as well as weak line features --- the only lines clearly visible are the \(1460.8\)~keV and \(2614.5\)~keV lines.
These features suggest that component~1 contains distant emission sources with appreciable attenuation and scattering.
The sharp rise in the continuum below \(400\)~keV is also suggestive of skyshine, the phenomenon where terrestrial emission is scattered down from the atmosphere above the detectors~\cite{hovgaard_reducing_1997}.

Component~2 contains the major \isot{Rn}{222}-series lines at \(242.0\), \(295.2\), \(351.9\), \(609.3\), \(1120.3\), and \(1764.5\)~keV as well as the positron annihilation line at \(511\)~keV\@.
It is the dominant component above \(2800\)~keV, where a continuum of photons due to cosmic-ray interactions makes up the majority of the background~\cite{sandness_accurate_2009}.
Thus component~2 seems to contain radon and cosmic emission.
The region around \(1460.8\) displays an artifact from the strong \isot{K}{40} line feature in both of the other components.

\begin{figure}[t!]
\centering
\includegraphics[width=0.9\columnwidth]{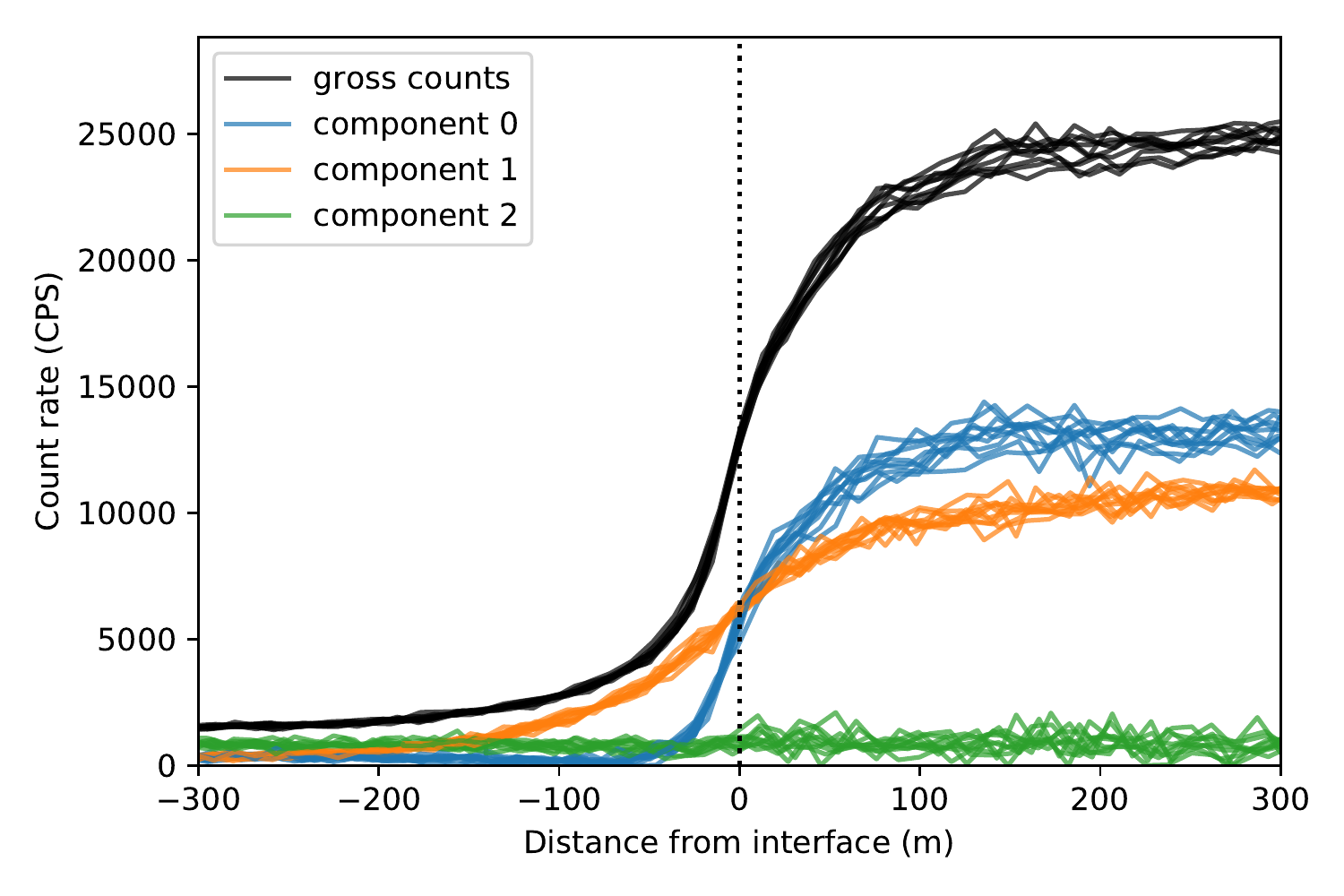}
\caption{NMF component weights during the ten crossings of the land-water interface in the Lake Mohave dataset.
Component~0 rises more rapidly than component~1 when approaching land, while component~2 is approximately constant.\label{fig:weights_interface}}
\end{figure}

The lower pane of \Fref{fig:lakemohave_decomp} shows the fitted weights for the three components during the entire dataset, and \Fref{fig:weights_interface} shows the same but only during the crossings of the land-water interface.
The behavior of the weights at the interface bolsters the previously mentioned spectroscopic interpretation of the NMF components.
As the aircraft approaches land, component~0 rises later and more rapidly than component~1, and component~2 stays relatively constant.

\subsection{Monte Carlo modeling of spectral components}\label{sec:lakemohave_modeling}
In order to quantitatively understand the spectral shapes of the NMF components, a Monte Carlo-based model of the detector system was used to generate spectral shapes for the postulated sources of background.
As in ref.~\cite{sandness_accurate_2009}, Monte Carlo simulations that can accurately model backgrounds require a geometry that extends hundreds of meters in all directions.
The authors of ref.~\cite{sandness_accurate_2009} modeled emission from the ground out to a range of \(300\)~m, and required an atmosphere that extended \(300\)~m above in order to accurately model skyshine from the ground.
Because our detector system is elevated above the ground, we extend the ground emission out to \(610\)~m (2000 ft) and allow air scattering up to a height and radius of \(5000\)~m.
Simulations of this scale are slow, so like ref.~\cite{sandness_accurate_2009} we make simplifying assumptions to break the simulation into smaller, more tractable parts.

For each part of the simulation, a response matrix was generated to relate the ``input'' spectrum to the ``output'' spectrum.
Each element \(R_{ij}\) of a response matrix \(\mathbf{R}\) gives the probability that an event in bin \(j\) of the input spectrum will result in an event in bin \(i\) of the output spectrum, so the output spectrum is found using a matrix dot product:
\begin{equation}
    \mathbf{x}^{\mathrm{out}} = \mathbf{R} \mathbf{x}^{\mathrm{in}}.
\end{equation}
For simplicity, the input flux distribution for all response matrices, unless otherwise noted, is proportional to the cosine of the normal to the relevant geometry's surface.

For terrestrial emission, we generated a response matrix \(\mathbf{R}_{\mathrm{terr}}\) for photon emission from within a slab of \(1.5\)~m thick soil by tallying the photon emission through the surface.
The soil composition used is the U.S. average from ref.~\cite{mcconn_jr_compendium_2011}.
The flux distribution that emerges is approximately proportional to the cosine of the normal to the slab surface, as assumed in the response matrices and as seen in ref.~\cite{sandness_accurate_2009}.

To simulate the transport of terrestrial photons through the atmosphere, a larger series of simulations were run using the MEGAlib Geant4-based Monte Carlo code~\cite{zoglauer_megalib_2006, agostinelli_geant4simulation_2003}.
These simulations involved monoenergetic gamma rays emitted from a point on a horizontal plane that represented the ground.
The photons were emitted into the upward direction with an angular distribution proportional to the cosine of the angle relative to the ground normal and were transported through dry sea-level air until absorption or they left a cylindrical volume of \(5000\)~m radius and height.
The tracking of photons that struck the ground surface were terminated.
Within the air, tally volumes defined by concentric cylinders were formulated to produce a series of track-length flux estimator tallies.
The photon energy and azimuth and elevation angles were also tracked in each tally volume, resulting in \(5^{\circ}\)-spaced angular bins and 31~energy bins consisting of 30 equally spaced down-scatter energy bins and a single full-energy bin.
The response due to a planar source is then inferred by leveraging symmetry and assuming that summing tallies at different radii is equivalent to integrating over the areal extent of the ground source.
These summed per-source photon fluxes (responses) were separately calculated at \(100\)~ft above a uniformly emitting ground plane for four geometric contributions to the overall flux: direct responses consisting of the flux from upward-directed photons emitted within (\(\mathbf{R}_{\mathrm{dir,near}}\)) or beyond (\(\mathbf{R}_{\mathrm{dir,dist}}\)) a ``cutoff'' radius \(r_{\mathrm{cutoff}}\) measured along the ground from the tally position, and skyshine responses consisting of the sum of all flux tally contributions from photons traveling downward emitted within (\(\mathbf{R}_{\mathrm{sky,near}}\)) or beyond (\(\mathbf{R}_{\mathrm{sky,dist}}\)) the cutoff radius.

For cosmic and radon backgrounds we simulated a \(3000 \times 3000 \times 1000\)~m rectangular prism of atmosphere and generated a response matrix for uniformly distributed isotropic emission within it and tallied photons that are incident on a \(1000 \times 1000\)~m window centered on the large face.
This atmospheric emission matrix was denoted \(\mathbf{R}_{\mathrm{atm}}\).
Radon and \(511\)~keV emission were propagated through this response matrix, but the cosmic continuum was not since it is an empirical model based upon detector measurements, not the physical production mechanism.

Finally, to simulate the response of the NaI detectors to all incident background sources, a response matrix \(\mathbf{R}_{\mathrm{abs}}\) was generated for photons incident on the large face of a \(2 \times 4 \times 16\)~inch NaI detector that are absorbed by the detector.

The complete response matrix for nearby emission is therefore
\begin{equation}
    \mathbf{R}_{\mathrm{near}} = \mathbf{R}_{\mathrm{abs}} (\mathbf{R}_{\mathrm{dir,near}} + \mathbf{R}_{\mathrm{sky,near}}) \mathbf{R}_{\mathrm{terr}},
\end{equation}
and the matrix for distant emission is
\begin{equation}
    \mathbf{R}_{\mathrm{dist}} = \mathbf{R}_{\mathrm{abs}} (\mathbf{R}_{\mathrm{dir,dist}} + \mathbf{R}_{\mathrm{sky,dist}}) \mathbf{R}_{\mathrm{terr}}.
\end{equation}
The complete matrix for radon and \(511\)~keV emission is
\begin{equation}
    \mathbf{R}_{\mathrm{Rn+511}} = \mathbf{R}_{\mathrm{abs}} \mathbf{R}_{\mathrm{atm}},
\end{equation}
while no response matrix was applied to the cosmic continuum.


\subsection{Results of Monte Carlo model fits}\label{sec:lakemohave_model_results}
To attribute the modeled background components to the NMF components, the following analysis was performed.
First, the NMF components were multiplied by their average weights to give them an absolute scale of counts per second per keV\@.
This scaling was performed so that when fitting near and distant KUT spectra to the components, not only their shapes but also their relative magnitudes could be constrained.
The variance of each element of this scaled \(\mathbf{V}\) matrix was estimated from the second derivatives of \(\Lambda \) with respect to the elements of \(\mathbf{V}\) (i.e., the Fisher information matrix).

The nine different postulated background components were then calculated for a given value of \(r_{\mathrm{cutoff}}\) and cosmic continuum power-law index:
\begin{itemize}
   \item Terrestrial \isot{K}{40} (nearby and distant)
   \item Terrestrial \isot{U}{238} series (nearby and distant)
   \item Terrestrial \isot{Th}{232} series (nearby and distant)
   \item Atmospheric \isot{Rn}{222} series
   \item Cosmic continuum
   \item Cosmic \(511\)~keV emission
\end{itemize}
For all sources except for the cosmic continuum, the input spectra \(\mathbf{x}^{\mathrm{in}}\) consisted of delta functions for each gamma-ray emission line weighted by branching ratio for the isotope or decay series assuming secular equilibrium.
The input spectrum for the cosmic continuum was modeled as a power law with arbitrary normalization.
All background components were then obtained by computing the dot product with the appropriate response matrix from the previous section.
Finally, all of the background components were convolved with a model of the detector energy resolution to produce realistic line widths.

In order to establish which background components contribute most strongly to each NMF component, a \(\chi^2\) minimization was performed to fit a linear combination of all nine background components simultaneously to the three NMF components, for a total of 27~parameters.
To find the best cutoff radius and cosmic power-law index, the \(\chi^2\) minimization was performed for each value of \(r_{\mathrm{cutoff}}\) in a grid from \(50\)~m to \(150\)~m in \(5\)~m increments, and for values of the power-law index between~\(0.00\) and~\(1.20\) in increments of~\(0.05\).
All coefficients were constrained to be non-negative, and the coefficients between the nearby and distant components of each terrestrial type (e.g., nearby and distant terrestrial \isot{K}{40}) were constrained so that their ratio was within \(20\% \) of the ratio for uniform infinite plane emission.
For all fits, the region below \(200\)~keV was excluded because in that region materials surrounding the detectors begin to strongly influence the shape of the background and those materials have not been modeled here.

After performing these fits with only the \(20\% \) constraint between near and distant terrestrial emission, the best-fit model was consistent with the basic hypothesis about the nature of the NMF components.
Component~0 was dominated by the nearby KUT background components and \isot{Rn}{222}, which closely resembles the nearby \isot{U}{238} background component.
Component~1 was dominated by the distant KUT components and nearby \isot{U}{238}.
Finally, component~2 was dominated by \isot{Rn}{222}, the cosmic continuum, nearby \isot{U}{238}, and cosmogenic \(511\)~keV\@.
The nearby \isot{U}{238} and \isot{Rn}{222} series spectra are similar in shape, which could explain the appearance of nearby \isot{U}{238} in the fit to NMF component~2.
The best value of \(r_{\mathrm{cutoff}}\) was \(85\)~m, while the best cosmic power-law index was~\(0.55\).

The fits were performed again but with the additional constraints of allowing only nearby KUT to fit component~0, distant KUT for component~1, and radon and cosmics for component~2.
The best overall fit was found when \(r_{\mathrm{cutoff}} = 85\)~m and the cosmic power-law index was~\(0.65\).
The best-fit cosmic power-law index was significantly lower than the value of~\(1.3\) measured in ref.~\cite{sandness_accurate_2009}.
Specific details about the different detectors may account for some of the difference, however such differences are unlikely to account for all of the discrepancy.
The different elevations above sea level, as well as the altitude above ground level, may also affect the power law observed.
Differences in how radon was modeled could also contribute.
The nature of this discrepancy remains unknown.

\begin{figure*}[t!]
\centering
\includegraphics[width=0.42\textwidth]{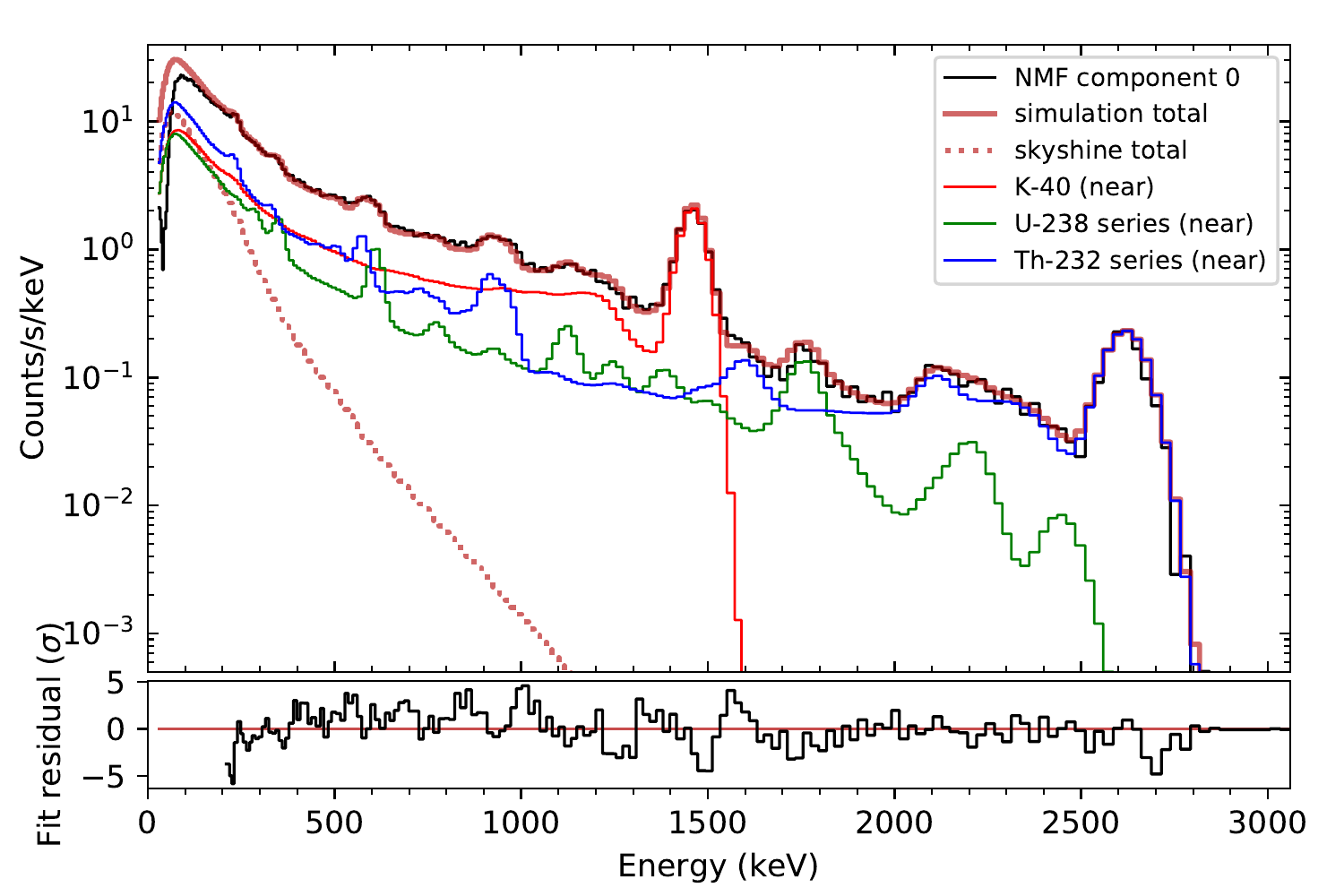}
\includegraphics[width=0.42\textwidth]{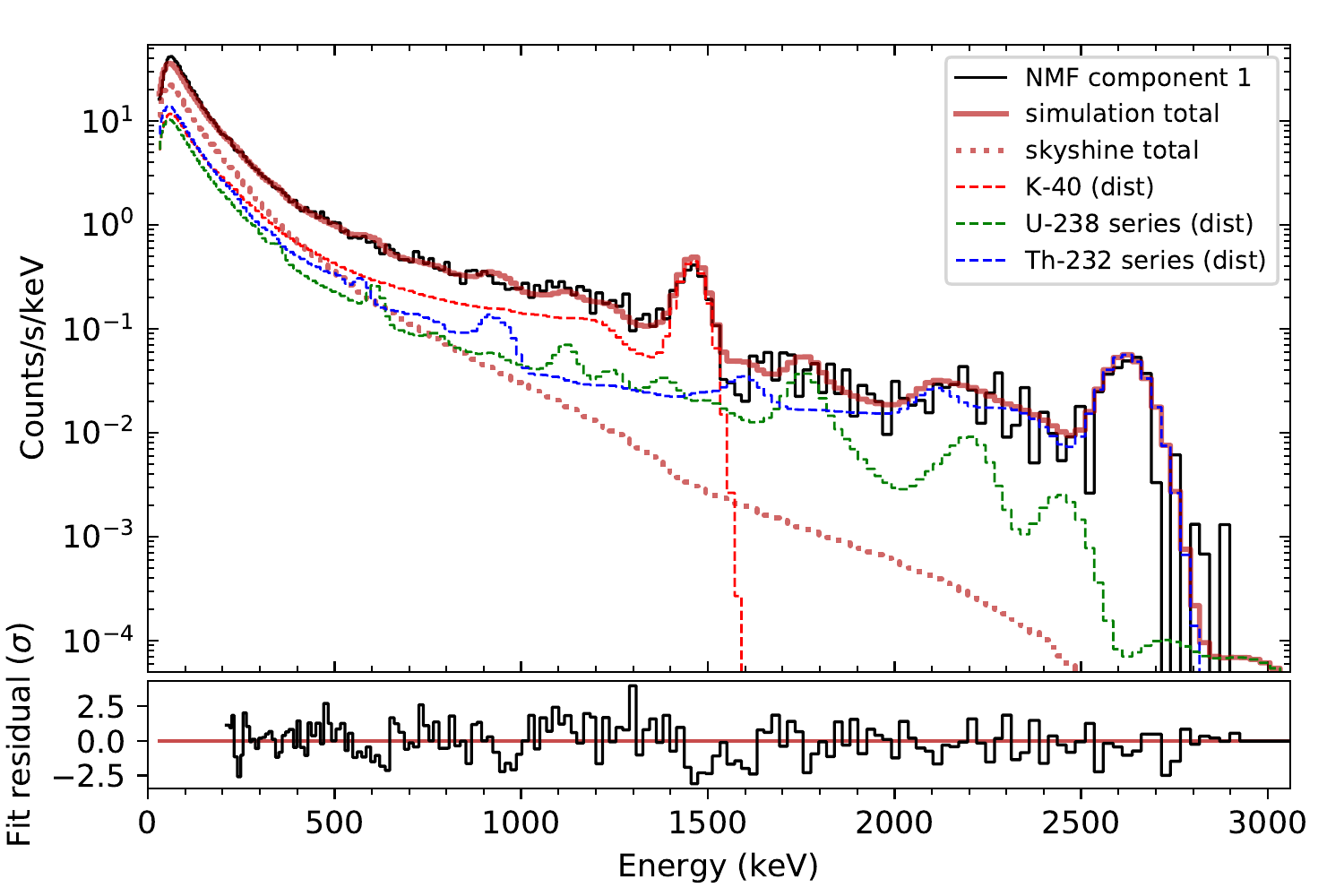}\\
\includegraphics[width=0.42\textwidth]{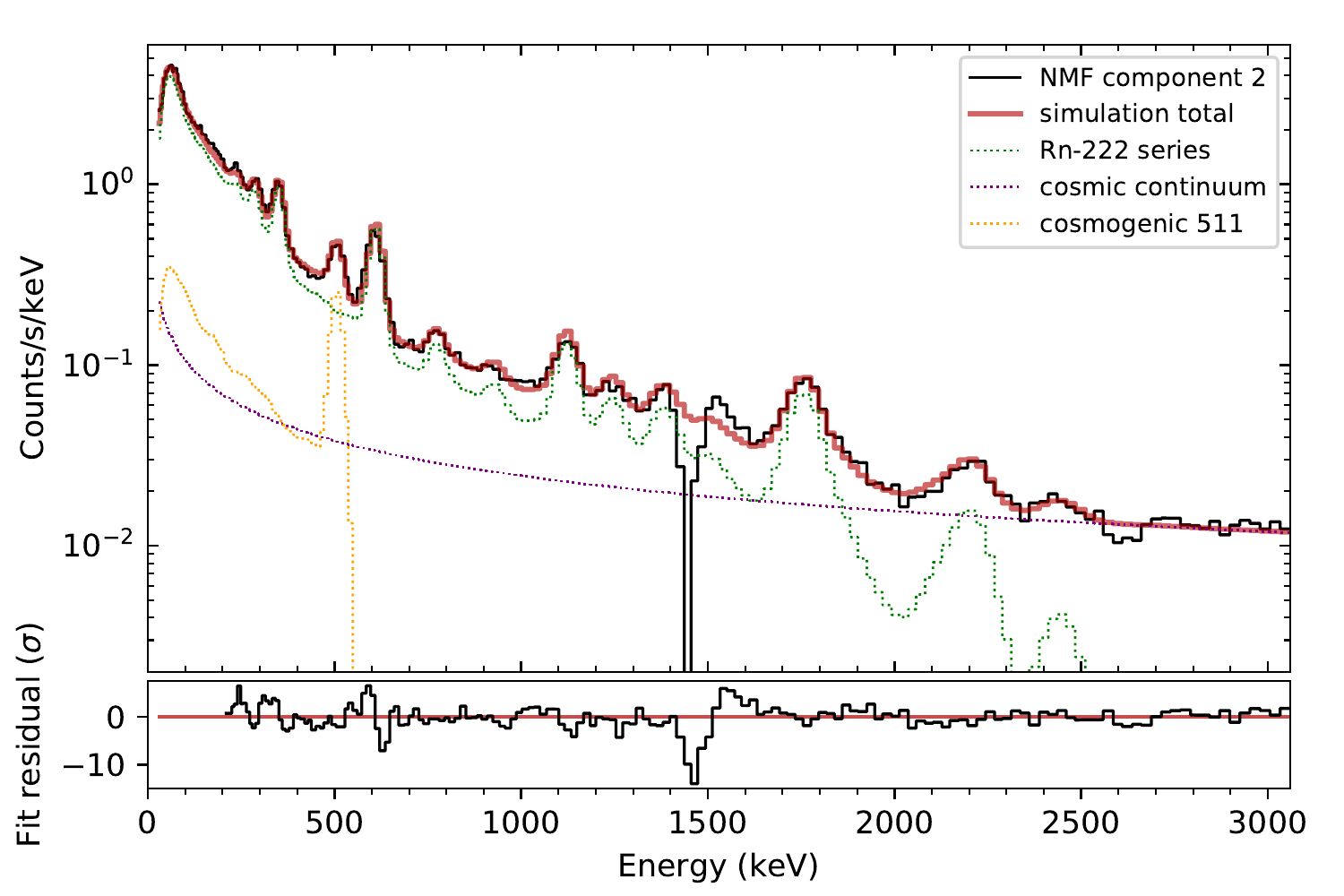}
\includegraphics[width=0.42\textwidth]{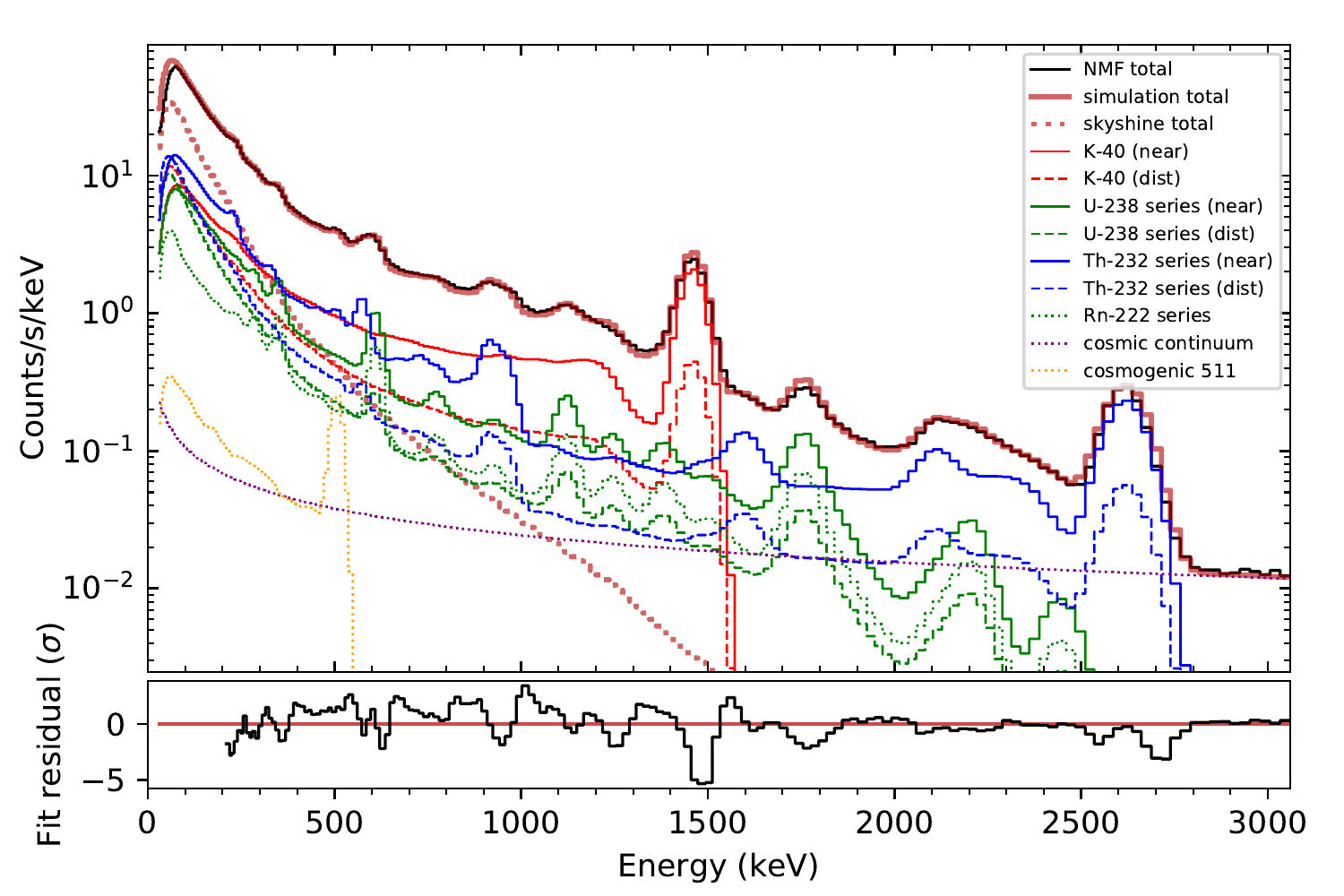}
\caption{Monte Carlo background components fit to the three NMF components (top left, top right, and bottom left) and the sum of all three components (bottom right) from the Lake Mohave dataset.
These fits provide strong evidence for the identification of component~0 with nearby emission, component~1 with distant emission, and component~2 with radon and cosmic emission.
The skyshine contributions from the terrestrial sources are shown separately.\label{fig:lakemohave_fits}}
\end{figure*}

\Fref{fig:lakemohave_fits} shows the results of these constrained fits.
The skyshine portion from all terrestrial components was summed together and plotted separately to show the portion of the component that is estimated to come from skyshine.
For component~0, the skyshine contribution makes up a maximum of about \(40\%\) of the spectrum below 100~keV, while for component~1 skyshine makes up two-thirds of events below \(100\)~keV and more than half of all events below \(400\)~keV\@.
Component~2 is dominated by radon, with smaller contributions from the cosmic continuum and \(511\)~keV emission.
The sum of all the weighted NMF components, which is approximately the mean measured spectrum, is also shown in \Fref{fig:lakemohave_fits}, together with the sums of all the fitted background components.

The NMF components and the background model fits are in general agreement and provide strong evidence for the identification of component~0 with nearby emission, component~1 with distant emission, and component~2 with radon and cosmic emission.


\subsection{Comparison to NASVD}\label{sec:nasvd}
Finally, we compare the NMF components to components derived using NASVD\@.
To perform NASVD, one scales each element of \(\mathbf{X}\) to have unit variance and then performs singular value decomposition (SVD) on the resulting matrix~\cite{hovgaard_new_1997}.
To estimate the mean of each bin, NASVD assumes that the average spectral shape is approximately the same.
Defining~\(\mathbf{s}\) to be the average of the rows of \(\mathbf{X}\) that is then normalized to sum to~\(1\) (i.e., the average unit-normalized spectrum):
\begin{equation}
    s_j \equiv \frac{\sum_i X_{ij}}{\sum_i \sum_j X_{ij}},
\end{equation}
and also defining the sum of the counts in each spectrum (i.e., the gross counts) to be~\(\mathbf{c}\):
\begin{equation}
    c_i \equiv \sum_j X_{ij},
\end{equation}
NASVD uses \(c_i s_j\) as an approximation to the mean of measurement \(X_{ij}\) and forms the unit-variance matrix \(\mathbf{X'}\):
\begin{equation}\label{eq:xprime}
    \mathbf{X'} \equiv \mathrm{diag}(\mathbf{c})^{-1/2}\, \mathbf{X}\, \mathrm{diag}(\mathbf{s})^{-1/2}.
\end{equation}
Then SVD is used to decompose \(\mathbf{X'}\):
\begin{equation}\label{eq:svd}
    \mathbf{X'} = \mathbf{U} \mathbf{\Sigma} \mathbf{W}^T,
\end{equation}
where \(\mathbf{U}\) is an \(n \times n\) unitary matrix, \(\mathbf{\Sigma}\) is an \(n \times d\) diagonal matrix of the singular values, and \(\mathbf{W}\) is a \(d \times d\) unitary matrix.
To remove noise, the highest \(k\) singular values are chosen and the rest are culled to form the \(n \times k\) matrix \(\mathbf{\Sigma}_k\) and the \(d \times k\) matrix \(\mathbf{W}_k\).
Now defining
\begin{eqnarray}
    \mathbf{A} &\equiv& \mathrm{diag}(\mathbf{c})^{1/2} \mathbf{U} \mathbf{\Sigma}_k \\
    \mathbf{V} &\equiv& \mathbf{W}_k^T \mathrm{diag}(\mathbf{s})^{1/2}
\end{eqnarray}
we obtain the NASVD solution to~\fref{eq:linear_decomp}.
A consequence of NASVD is that the first column of \(\mathbf{U} \mathbf{\Sigma}_k\) is~\(\sqrt{\mathbf{c}}\) and the first column of \(\mathbf{W}_k\) is~\(\sqrt{\mathbf{s}}\), so therefore the first column of \(\mathbf{A}\) is~\(\mathbf{c}\) and the first row of \(\mathbf{V}\) is~\(\mathbf{s}\)~\cite{hovgaard_airborne_1998}.
The remaining components are additive perturbations to the mean spectrum in order of decreasing variance, and the row-wise sums of \(\mathbf{V}\) beyond the first component are zero.
(This latter fact is a result of the orthogonality of \(\mathbf{W}_k\).
Letting \(\mathbf{w}_j\) be the \(j\)th column of \(\mathbf{W}_k\), then since the first column of \(\mathbf{W}_k\) is \(\mathbf{w}_0 = \sqrt{\mathbf{s}}\), the sum of the \(j\)th row of \(\mathbf{V}\) is \(\mathbf{w}_j \cdot \sqrt{\mathbf{s}} = \mathbf{w}_j \cdot \mathbf{w}_0 = \delta_{0j}\).)

\begin{figure*}[t!]
\centering
\includegraphics[width=0.42\textwidth]{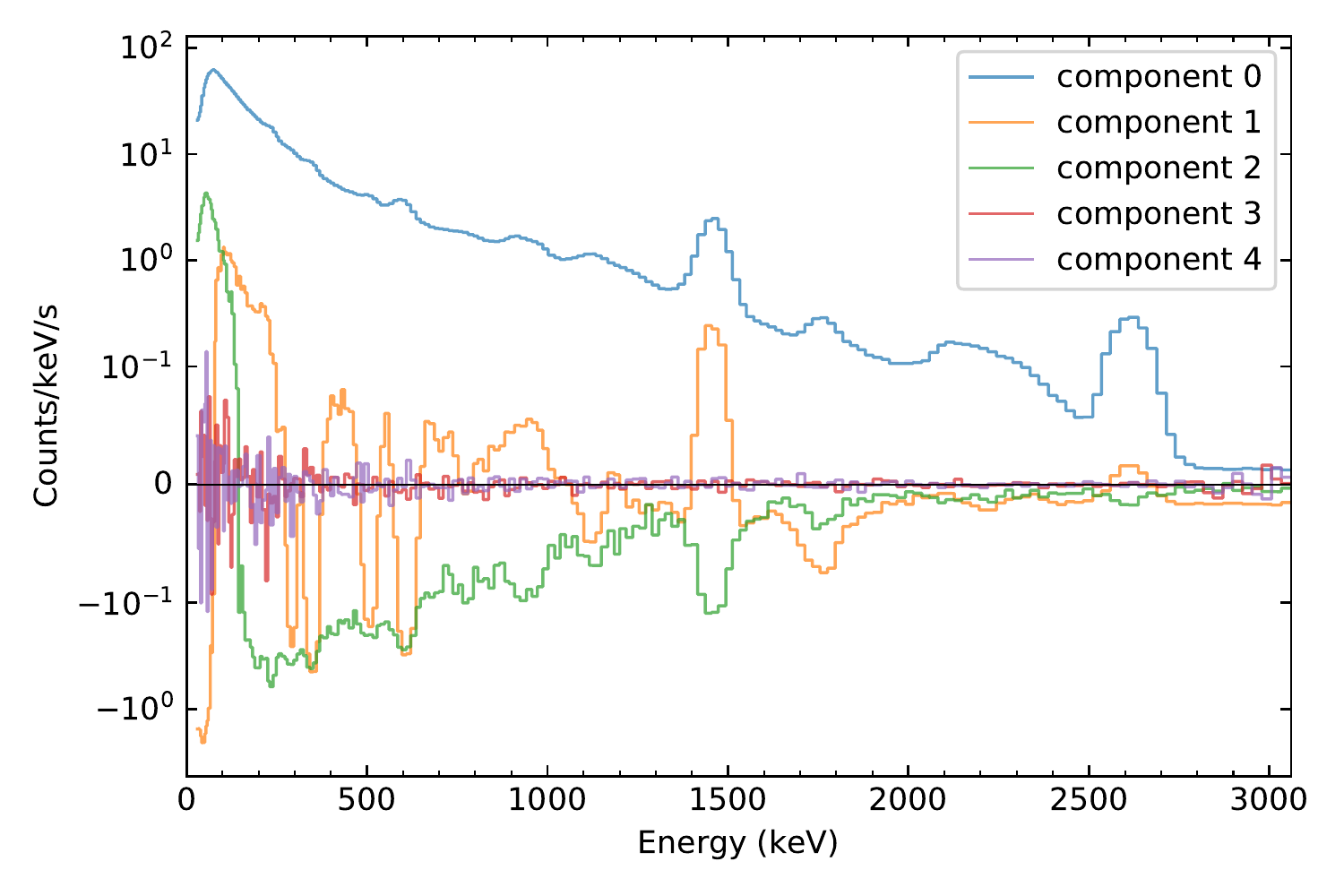}\\
\includegraphics[width=0.85\textwidth]{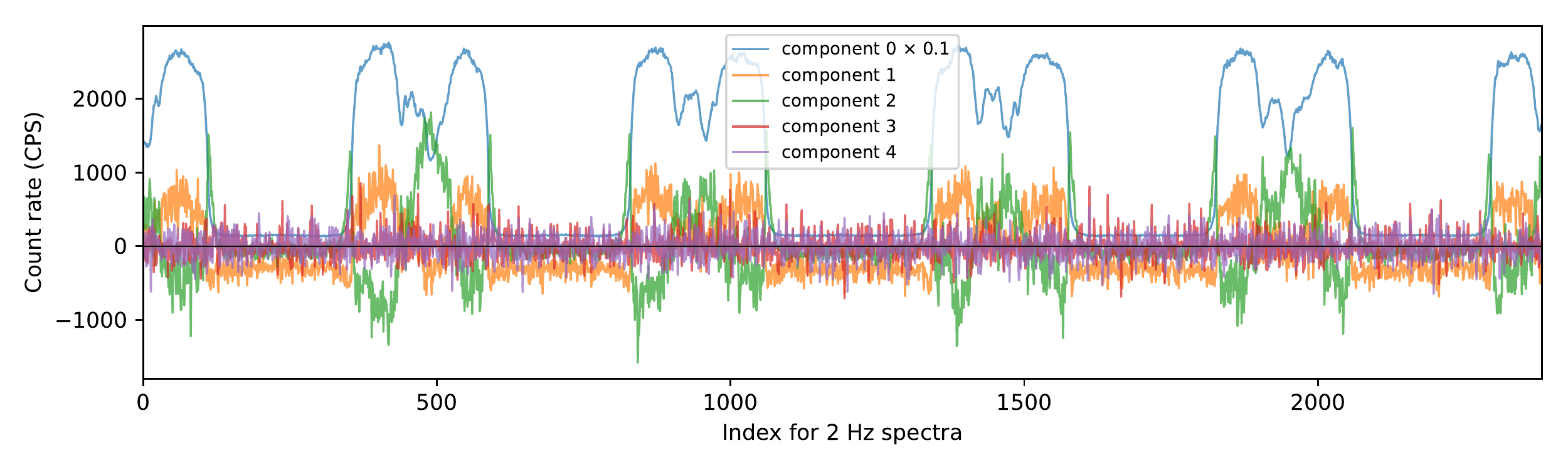}
\caption{NASVD components for the Lake Mohave data (top) and component weights (bottom).
Spectral components have been multiplied by their maximum weight to show their scale relative to each other.
In the bottom plot, the component~0 weights have been scaled by a factor of~\(0.1\) for ease of comparison to the others.
The first NASVD component is always the average spectral shape and its weights are the gross counts, while the following components represent variations from the average shape.
Components~0--2 account for over \(95\%\) of the variance of the dataset while components~3 and higher largely contain statistical noise.\label{fig:lakemohave_nasvd}}
\end{figure*}

\begin{figure}[t!]
\centering
\includegraphics[width=0.9\columnwidth]{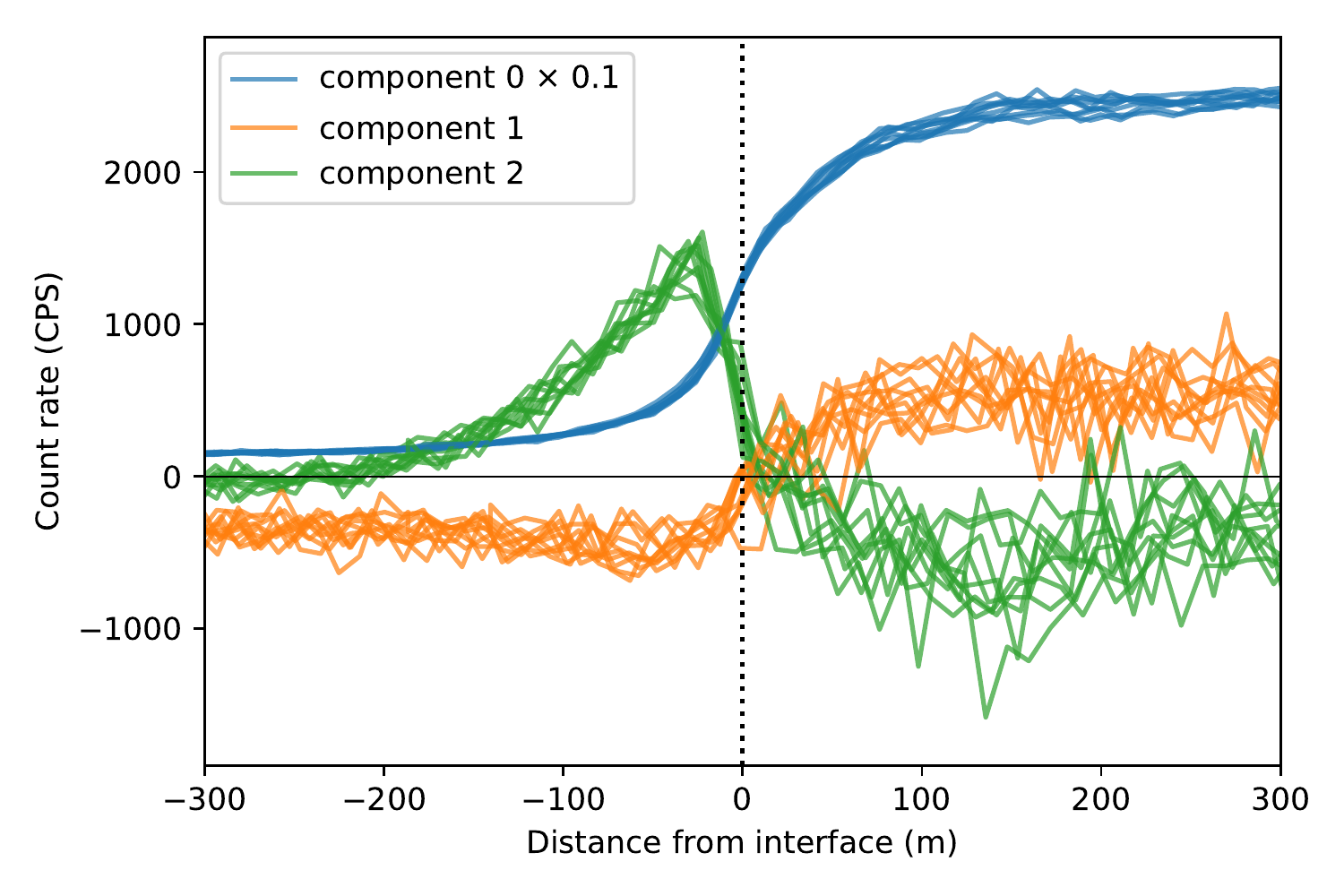}
\caption{Weights of the first three NASVD components during the ten crossings of the land-water interface in the Lake Mohave dataset (cf. the NMF weights in~\Fref{fig:weights_interface}).
The weights for component~0 are the gross counts, while components~1 and 2 change the relative importance of radon and cosmics and distant emission, respectively.
The component~0 weights have been scaled by a factor of~\(0.1\) for ease of comparison to the others.\label{fig:weights_interface_nasvd}}
\end{figure}

\Fref{fig:lakemohave_nasvd} shows the results of an NASVD decomposition for the Lake Mohave dataset.
Components~0--2 account for over \(95\%\) of the variance, while components~3 and higher have much smaller variances than the first three and do not show coherent spectral features.
Components~1 and 2 clearly display features that can be associated with spectroscopic phenomena, such as photopeaks, but because they take on negative values, each component cannot be the direct result of background source emission.
For example, increasing the weight of component~1 decreases the relative contribution of radon and cosmics while increasing the contribution of nearby \isot{K}{40} and \isot{Tl}{208} emission.
Likewise, increasing the weight of component~2 increases the contribution from distant emission relative to nearby emission.
As a further comparison to the NMF decomposition, \Fref{fig:weights_interface_nasvd} shows the weights of the first three NASVD components during the ten crossings of the land-water interface and is the equivalent of \Fref{fig:weights_interface}.
At the interface, the weights for components~1 and 2 change sign to account for the rapidly changing shape of the spectrum in that region.

In order to quantitatively compare the abilities of NASVD and NMF to represent the same spectra, we first note that the gross counts of both NASVD- and NMF-reconstructed spectra exactly match the gross counts of the measured spectra.
For NASVD, this property follows from the fact that the first column of \(\mathbf{A}\) are the gross counts \(\mathbf{c}\), and the sum of the \(j\)th row of \(\mathbf{V}\) is \(\delta_{j0}\).
For NMF, this property follows from the fact that when maximizing Poisson likelihood, any model that has an effective scale factor will be fit so that the sum of the model matches the measured gross counts.

\begin{figure}[t!]
\centering
\includegraphics[width=0.9\columnwidth]{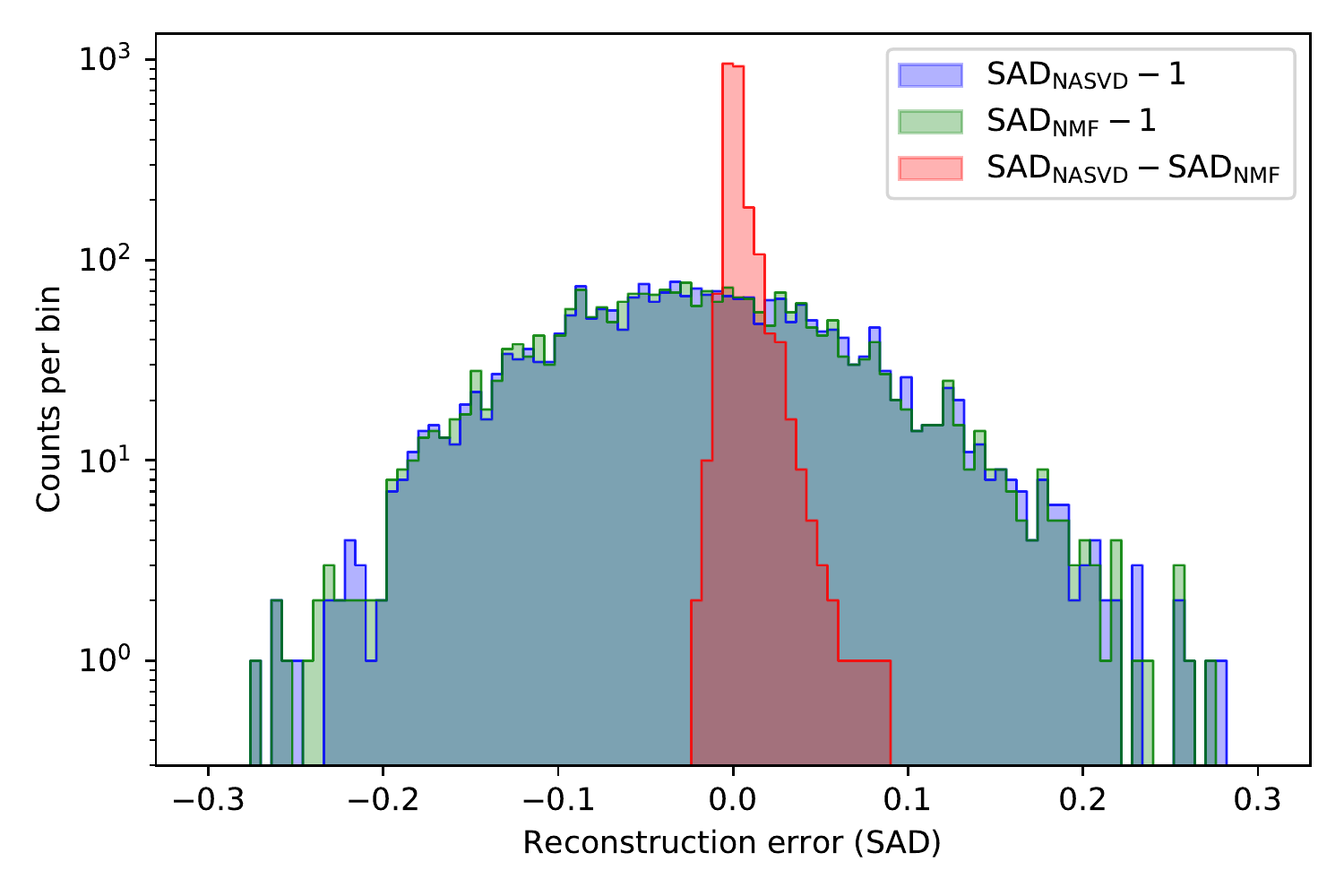}
\caption{Spectral Anomaly Detection (SAD) metric for spectra reconstructed using NASVD and NMF with 3~components.
The red histogram shows the difference in the metric between the two methods.
This difference is small, indicating that they reconstruct the measured spectra with similar fidelity.\label{fig:sad}}
\end{figure}

Since both methods exactly reconstruct each spectrum's gross counts, it is also important to compare how well they fit the shape of the spectrum.
For this purpose we introduce the Spectral Anomaly Detection (SAD) metric, the L2 norm between the \(i\)th measured spectrum and the \(i\)th reconstructed spectrum~\cite{miller_gamma-ray_2018, bilton_non-negative_2019}:
\begin{equation}
    \mathrm{SAD}_i \equiv \frac{\sum_j || X_{ij} - \hat{X}_{ij} ||^2}{\sum_j || X_{ij} ||},
\end{equation}
where the denominator normalizes the metric to a mean of unity.
We evaluated SAD for NMF and NASVD reconstructions with three components and histogrammed the results in~\Fref{fig:sad}.
The distribution of the difference in the SAD metric between the NASVD and NMF reconstructions is much narrower than the distribution of either SAD metric, indicating that the two methods reconstruct the measured spectra with similar fidelity.
This result is not surprising since both models have the same linear form with similar numbers of free parameters, though NMF has additional (non-negativity) constraints.
Notably, the histogram of SAD metric differences has an average value greater than zero, which indicates that NMF yields slightly better reconstructions than NASVD, at least according to the SAD metric.
The average value of the SAD metric difference becomes negative for this dataset (i.e., favoring NASVD) only once the number of NASVD components used in the reconstruction has been increased to nine.

\section{Application to San Francisco Bay Area surveys}\label{sec:bayarea}
To demonstrate the application of NMF decomposition under more general conditions, two survey datasets from the San Francisco Bay Area (henceforth, Bay Area) were analyzed using the techniques of the previous section.
These datasets were part of a larger survey campaign performed in August~2012~\cite{nstec_remote_sensing_laboratory_aerial_2012}.
The first set was taken at the Port of Oakland on 30~August 2012 and consisted of 5,687 \(1\)-Hz spectra, and the second dataset was collected in Pacifica on 28~August 2012 and consisted of 4,505 \(1\)-Hz spectra.
The data were taken with the same NaI(Tl) detector system as the Lake Mohave data, and the same energy range and binning scheme were used.
Both the data rate (\(1\)~Hz) and the nominal flight altitude above ground level (\(300\)~ft) differed from the Lake Mohave dataset (\(2\)~Hz and \(100\)~ft, respectively).
Both surveys contain land-water interfaces like the Lake Mohave survey, but both also contain roads and man-made structures.
The Pacifica survey is also of special interest because it contains rugged terrain and some geological features with substantially lower potassium content than typical (and therefore weak \isot{K}{40} emission), providing additional spectral variability.


\subsection{Anomaly detection and removal}\label{sec:bayarea_anomalies}
For the Bay Area datasets, the Spectral Comparison Ratio Anomaly Detection (SCRAD) method~\cite{pfund_examination_2007, jarman_comparison_2008, detwiler_spectral_2015} was applied to the data to identify and remove any spectral anomalies that could affect the NMF decomposition.
To use SCRAD one needs to bin each spectrum into a series of predetermined bins, track estimates of the means and covariances of the bins over time using an exponentially weighted moving average (EWMA), and then construct a vector of spectral comparison ratios (SCRs) from the binned counts.
SCRAD relies on the background estimate being accurate (we used an EWMA with a parameter of~\(0.1\)) and each bin being large enough that its statistics are Gaussian, not Poisson.

For SCRAD, 13~non-overlapping spectral windows were used with boundaries of \(30\), \(60\), \(80\), \(100\), \(120\), \(140\), \(160\), \(200\), \(250\), \(300\), \(400\), \(700\), \(1200\), and \(3060\)~keV\@.
These windows were chosen to span the entire energy range and so that no counts in any bin were fewer than~\(30\) in order to ensure the Gaussian approximation was statistically valid.
No effort was made to optimize the windows for any particular types of anomalies, nor was nuisance rejection implemented (i.e., N-SCRAD~\cite{pfund_improvements_2016}).
The SCR distance metric \(D_{SCR}^2\) was used as the test statistic~\cite{jarman_comparison_2008}, and its expected chi-squared distribution with \(12\)~degrees of freedom was used to set a threshold of \(5\sigma \) significance.

Using this metric, three anomalies were found in the Port of Oakland dataset.
These anomalies, at indices \(1964\), \(2304\), and \(4040\), can be seen in \Fref{fig:bayarea_weights} where they are marked with red arrows.
The anomaly at index \(1964\) lasted at least \(6\)~seconds and is of unknown origin.
It consisted of a hard continuum extending beyond \(3\)~MeV.
The anomaly at index \(2304\) consisted of an increase in counts around \(100\)~keV and occurred when the aircraft was near the position of the first anomaly on the following flight line, so it is likely down-scattered photons from the first anomaly.
The third anomaly at index \(4040\) was likely a \isot{Tc}{99\mathrm{m}} source due to elevated counts up to \(140\)~keV\@.
The data associated with these anomalies were excluded from the NMF training and are not shown on the maps in \Fref{fig:bayarea_maps}.


\subsection{Results of NMF decompositions}\label{sec:bayarea_decomp}
\begin{figure*}[t!]
    \centering
    \includegraphics[width=0.42\textwidth]{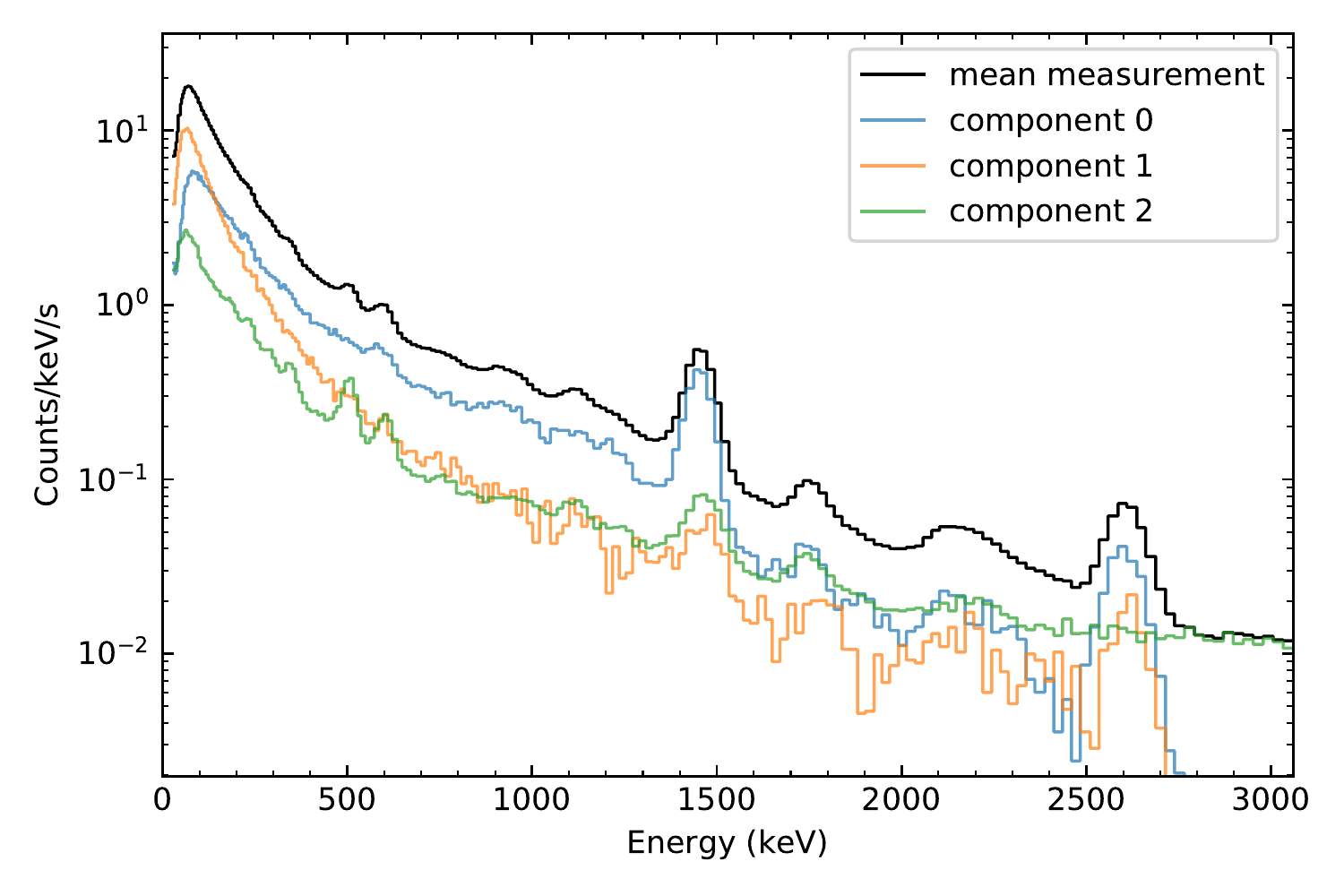}
    \begin{tikzpicture}
        \node[anchor=south west,inner sep=0] (image) at (0,0) {\includegraphics[width=0.42\textwidth]{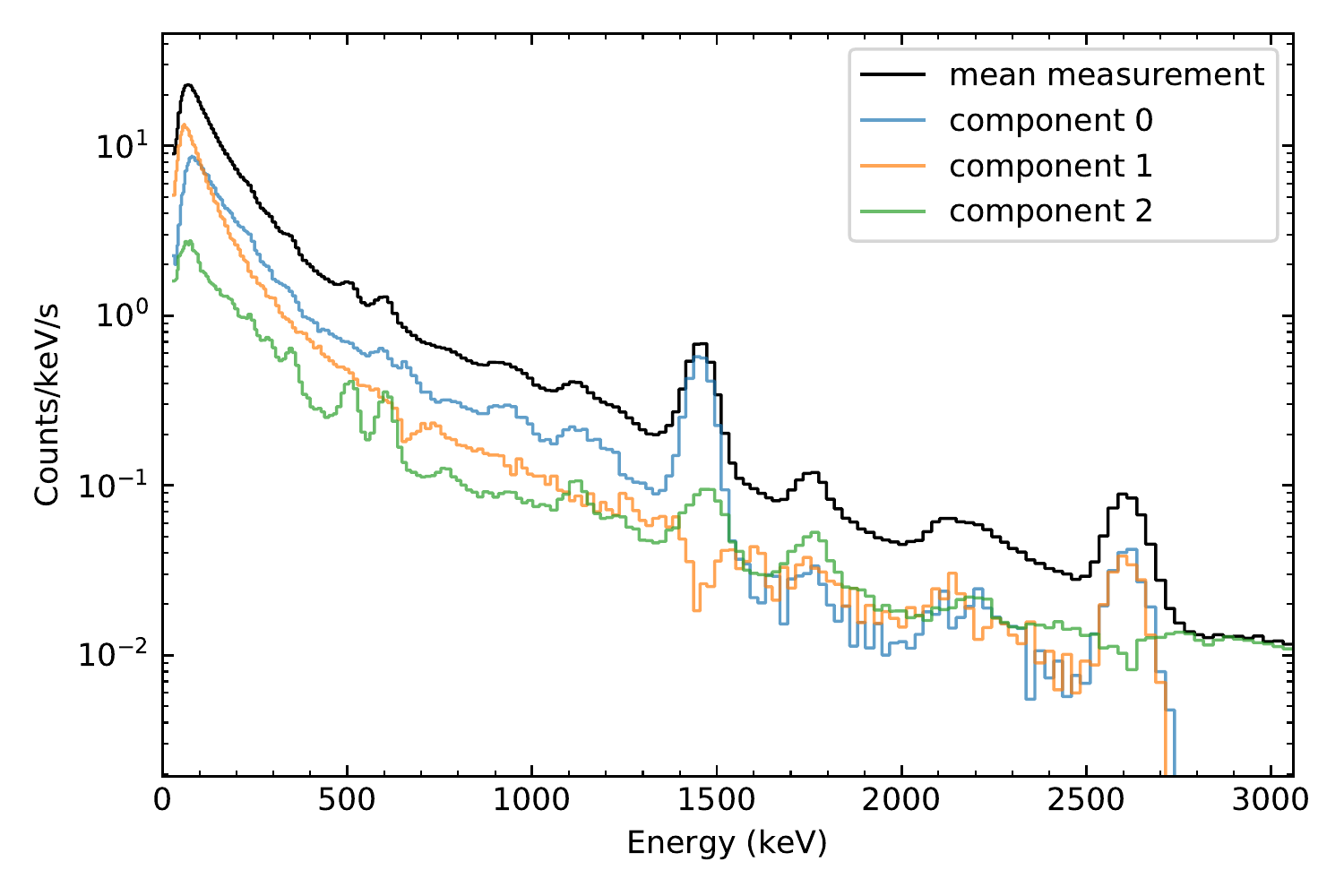}};
        \begin{scope}[x={(image.south east)},y={(image.north west)}]
            \draw [line width=1pt, red, opacity=0.7, -stealth]
                ({(74 + (662 / 3060) * 504) / 601}, 0.703) -- ++(0.0, -0.1);
        \end{scope}
    \end{tikzpicture}
\caption{NMF components for a three-component factorization fitted to the Port of Oakland data (left) and Pacifica (right).
These decompositions have a strong resemblance with each other and with the Lake Mohave decomposition (\Fref{fig:lakemohave_decomp}).
A red arrow marks the presence of the \(662\)~keV line from \isot{Cs}{137} in component~0 from Pacifica.\label{fig:bayarea_comps}}
\end{figure*}

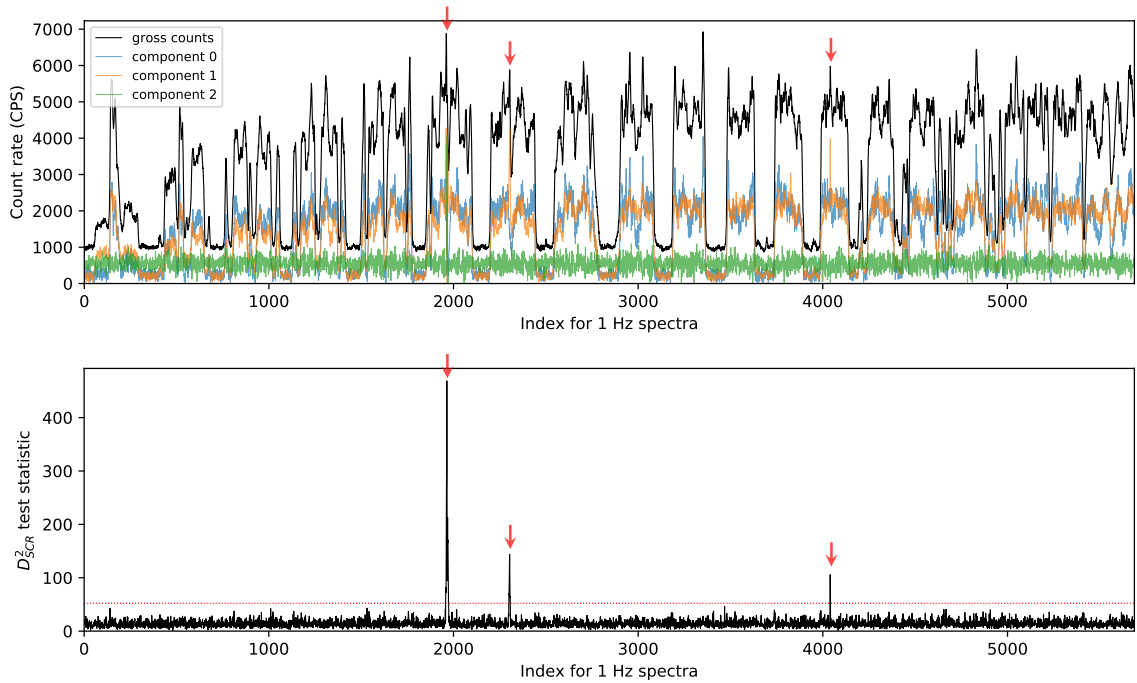
\begin{figure*}[t!]
    \centering
    \begin{tikzpicture}
        \node[anchor=south west,inner sep=0] (image) at (0,0) {\includegraphics[width=0.85\textwidth]{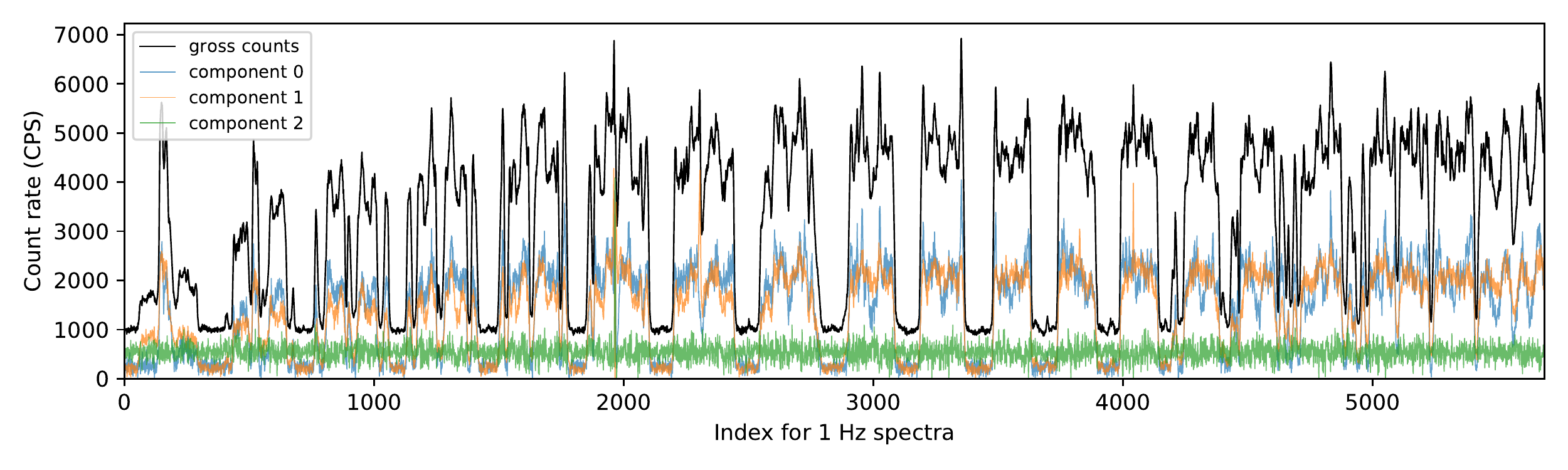}};
        \begin{scope}[x={(image.south east)},y={(image.north west)}]
            \draw [line width=1pt, red, opacity=0.7, -stealth]
                ({(145 + (1964 / 5678) * 1649) / 1823}, 0.99) -- ++(0.0, -0.07);
            \draw [line width=1pt, red, opacity=0.7, -stealth]
                ({(145 + (2304 / 5678) * 1649) / 1823}, 0.89) -- ++(0.0, -0.07);
            \draw [line width=1pt, red, opacity=0.7, -stealth]
                ({(145 + (4040 / 5678) * 1649) / 1823}, 0.90) -- ++(0.0, -0.07);
        \end{scope}
    \end{tikzpicture}
    \begin{tikzpicture}
        \node[anchor=south west,inner sep=0] (image) at (0 ,0) {\includegraphics[width=0.85\textwidth]{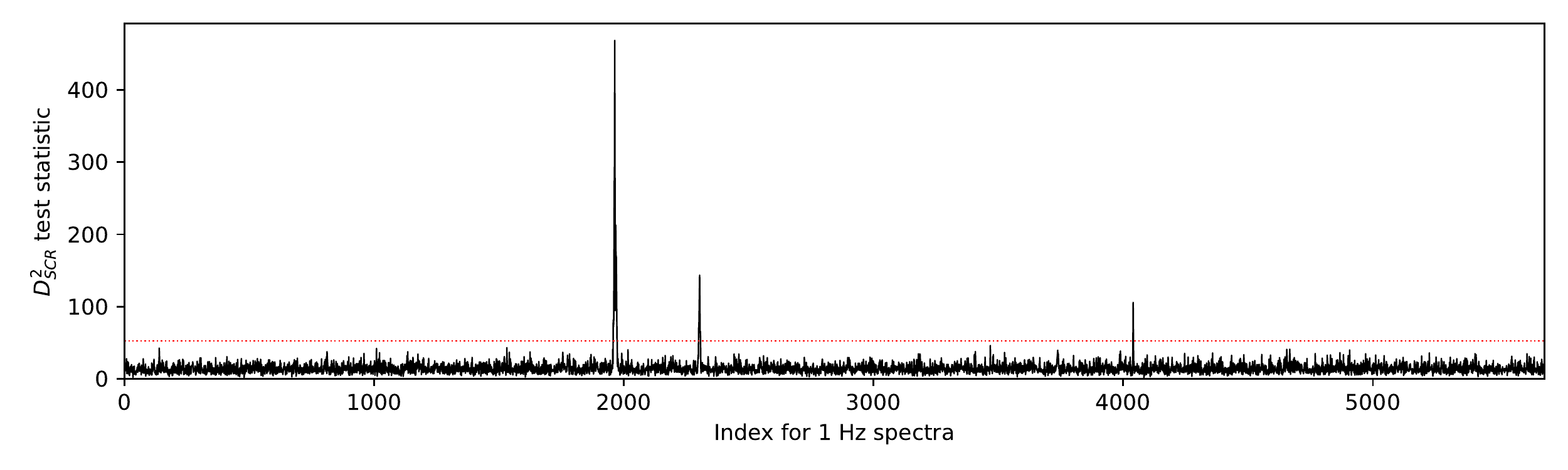}};
        \begin{scope}[x={(image.south east)},y={(image.north west)}]
            \draw [line width=1pt, red, opacity=0.7, -stealth]
                ({(145 + (1964 / 5678) * 1649) / 1822}, 0.99) -- ++(0.0, -0.07);
            \draw [line width=1pt, red, opacity=0.7, -stealth]
                ({(145 + (2304 / 5678) * 1649) / 1822}, 0.50) -- ++(0.0, -0.07);
            \draw [line width=1pt, red, opacity=0.7, -stealth]
                ({(145 + (4040 / 5678) * 1649) / 1822}, 0.45) -- ++(0.0, -0.07);
        \end{scope}
    \end{tikzpicture}
\caption{The gross count rate and NMF component weights for the Port of Oakland dataset (top).
The bottom plot shows the SCRAD test statistic used to identify the anomalies in the dataset, and the red line indicates a threshold of \(5\sigma\) significance.
The red arrows mark the locations of the anomalies.\label{fig:bayarea_weights}}
\end{figure*}

NMF was initialized and trained on the Bay Area datasets in the same manner as the Lake Mohave dataset.
NMF models with \(k=1\) to \(4\) components were trained, and AIC was again used to determine the optimal number of components to describe the dataset, resulting in \(k=3\) components for both the Port of Oakland and Pacifica datasets.
\Fref{fig:bayarea_comps} shows the resulting NMF components for each dataset, and \Fref{fig:bayarea_weights} shows the weights for the Port of Oakland dataset.

NMF decompositions are not expected to be identical in all situations since they are fitted to the training data provided.
For example, the Pacifica survey was over rugged terrain with regions of abnormally low \isot{K}{40}, while the Port of Oakland survey was over flat terrain but with man-made materials, which could give rise to sharp changes in KUT concentrations.
The datasets may also have different ambient backgrounds, which would affect any terms that are approximately constant.

In general both NMF decompositions are qualitatively similar to each other and to the Lake Mohave decomposition (\Fref{fig:lakemohave_decomp}), aside from some notable differences.
Component~0 on average makes up a smaller fraction of the total spectrum, which may be due to greater attenuation of nearby emission at the higher altitude.
One subtle difference in component~0 from Pacifica is that there is an indication of a weak \isot{Cs}{137} peak at \(662\)~keV from 1950s weapons-testing fallout, perhaps due to less surface disturbance than the Port of Oakland and greater deposition than Lake Mohave due to higher yearly rainfall.
In component~1, the \isot{K}{40} peak at \(1460.8\)~keV is less prominent (Oakland) or not at all present (Pacifica).
Component~2 in both Bay Area datasets now includes a clear \isot{K}{40} peak, which could be due to different aircraft background and the presence of potassium in the salt water of the Pacific Ocean and the San Francisco Bay.

\begin{figure*}[t!]
\centering
\includegraphics[width=0.48\textwidth]{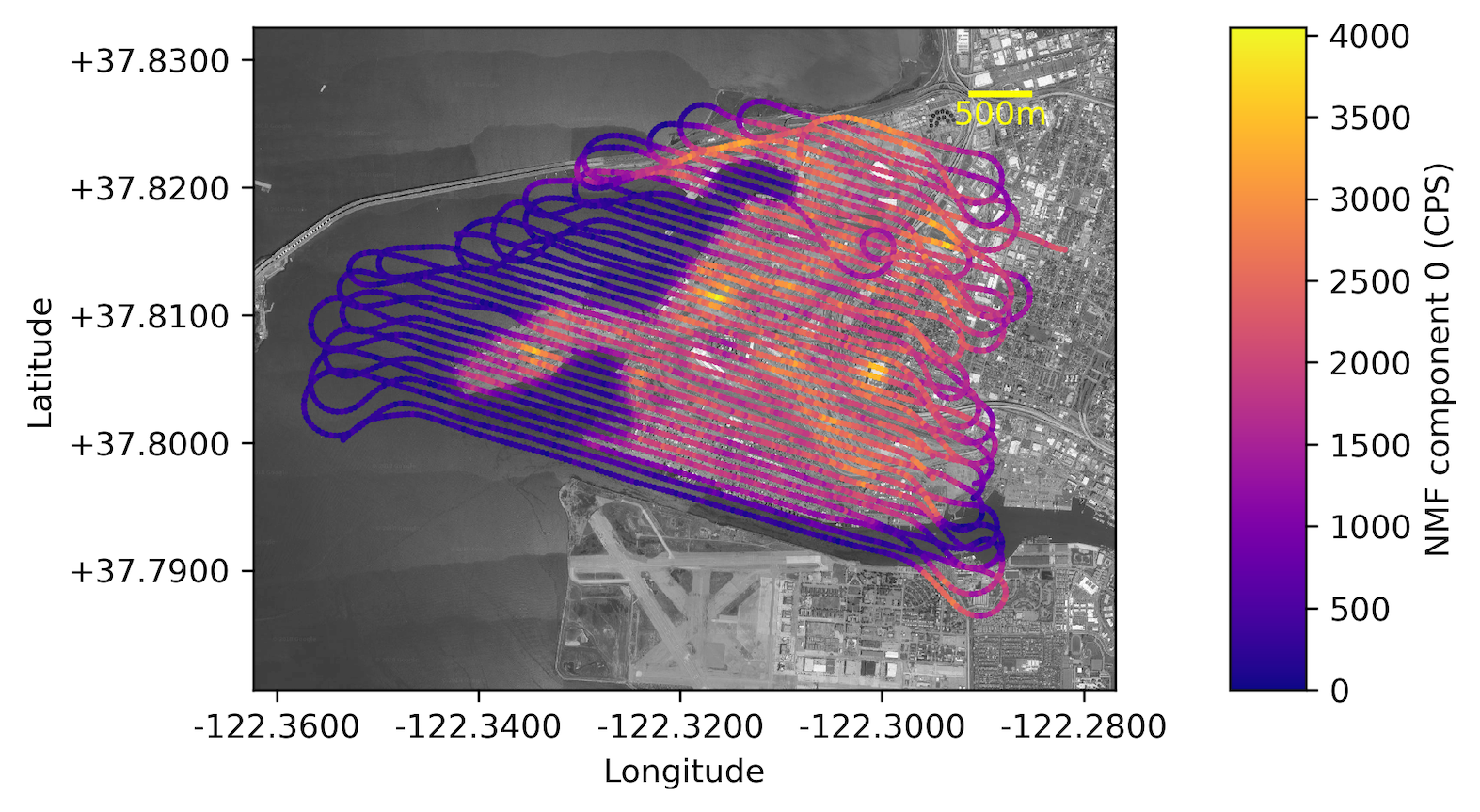}
\includegraphics[width=0.48\textwidth]{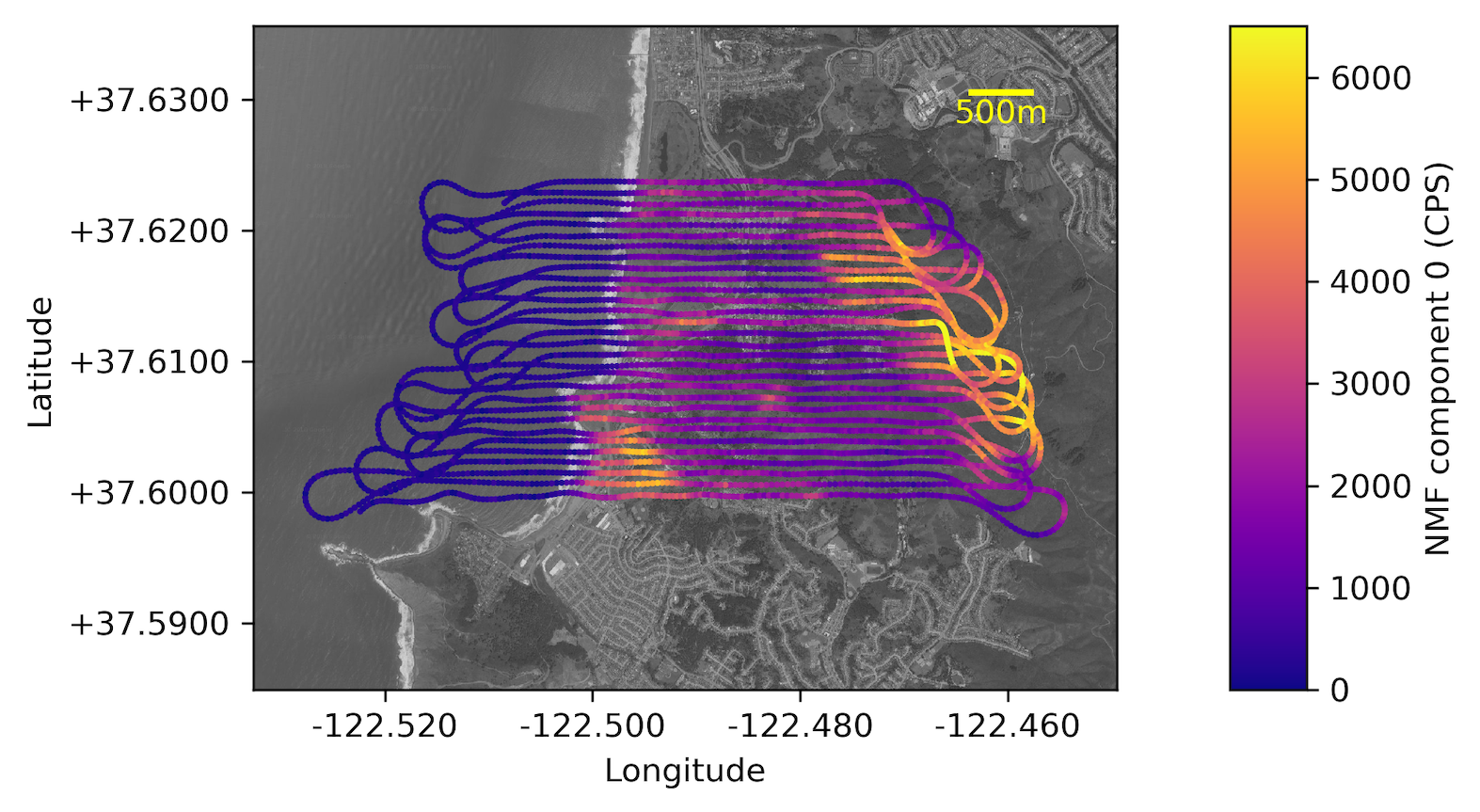}\\
\includegraphics[width=0.48\textwidth]{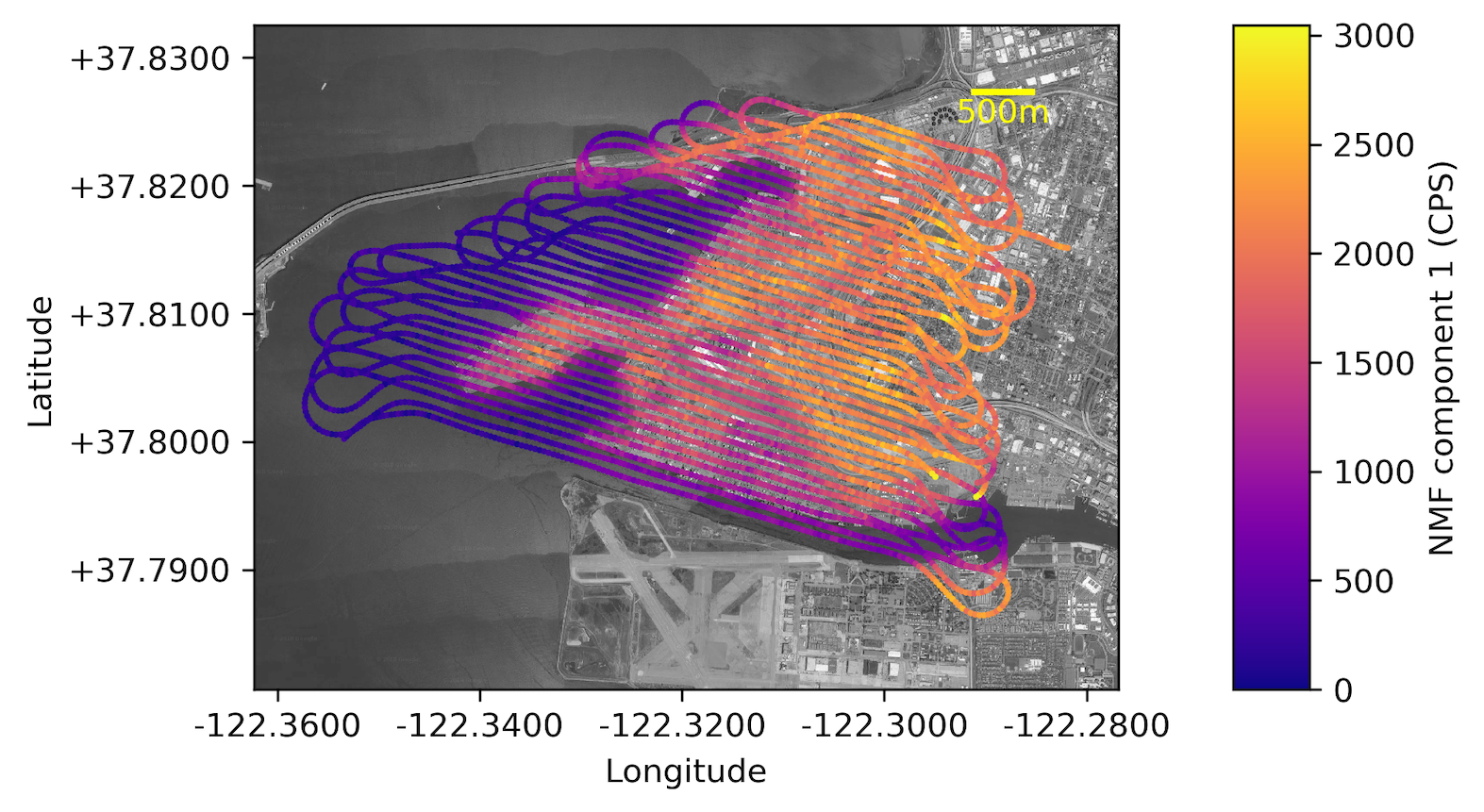}
\includegraphics[width=0.48\textwidth]{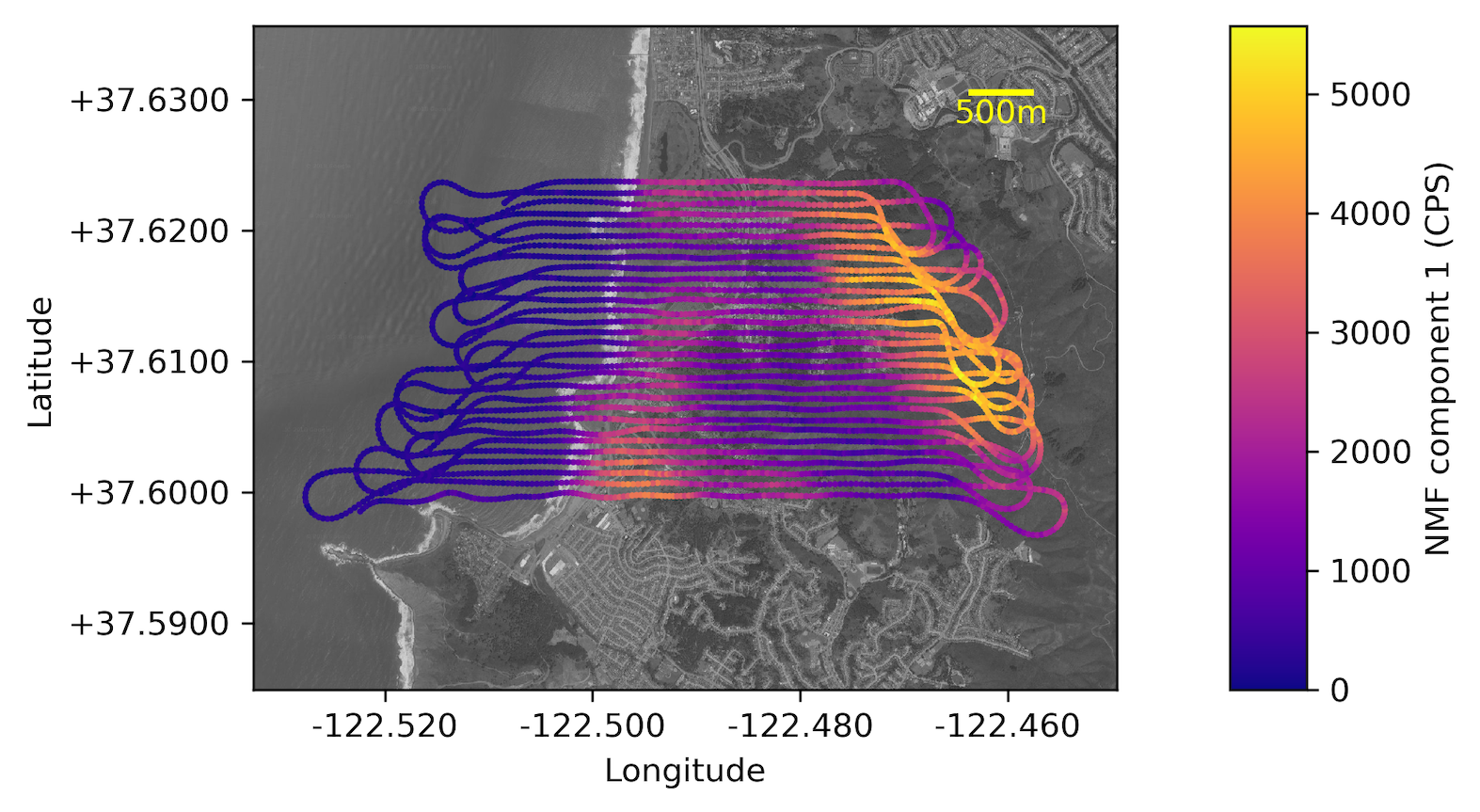}\\
\includegraphics[width=0.48\textwidth]{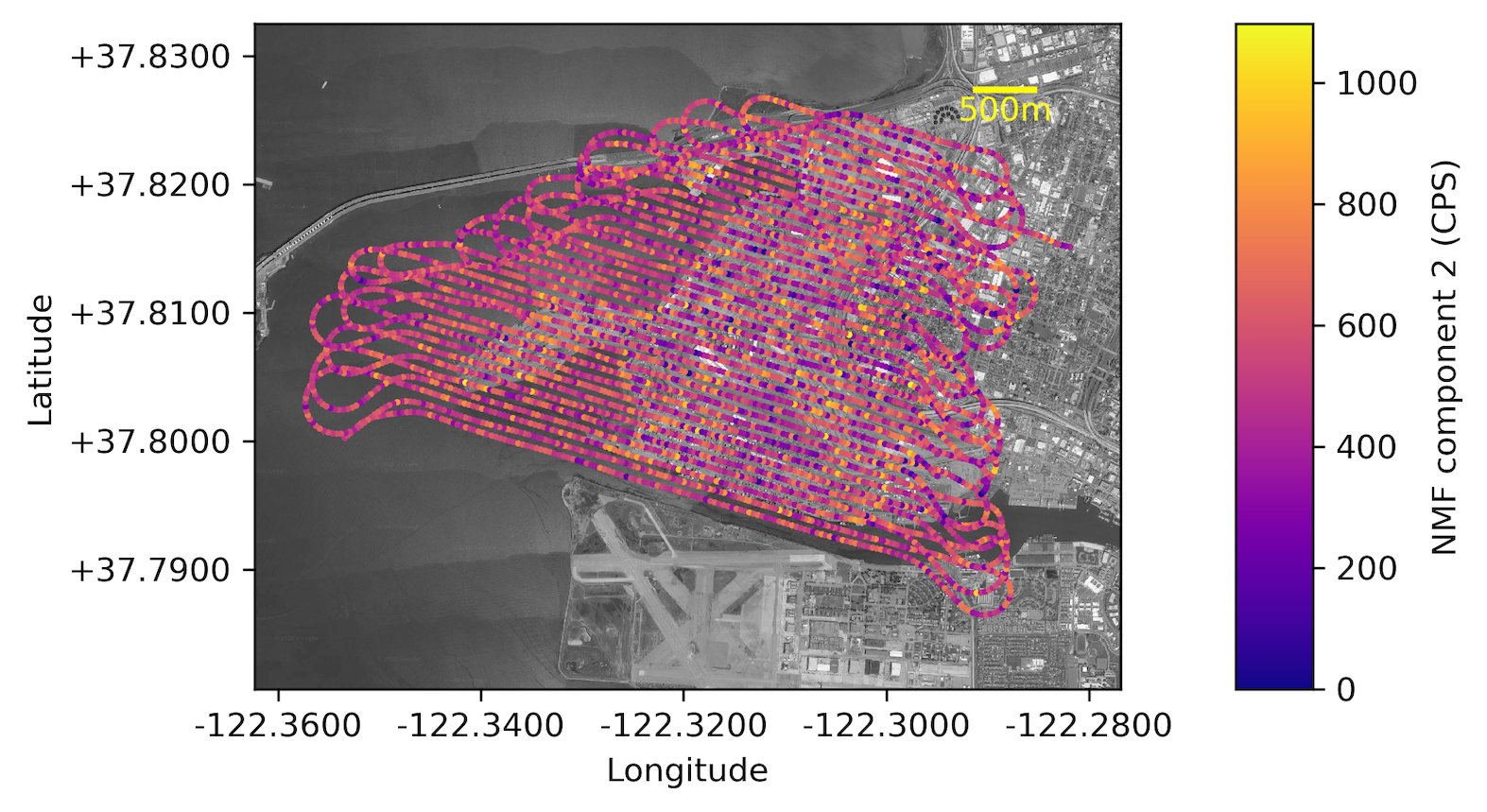}
\includegraphics[width=0.48\textwidth]{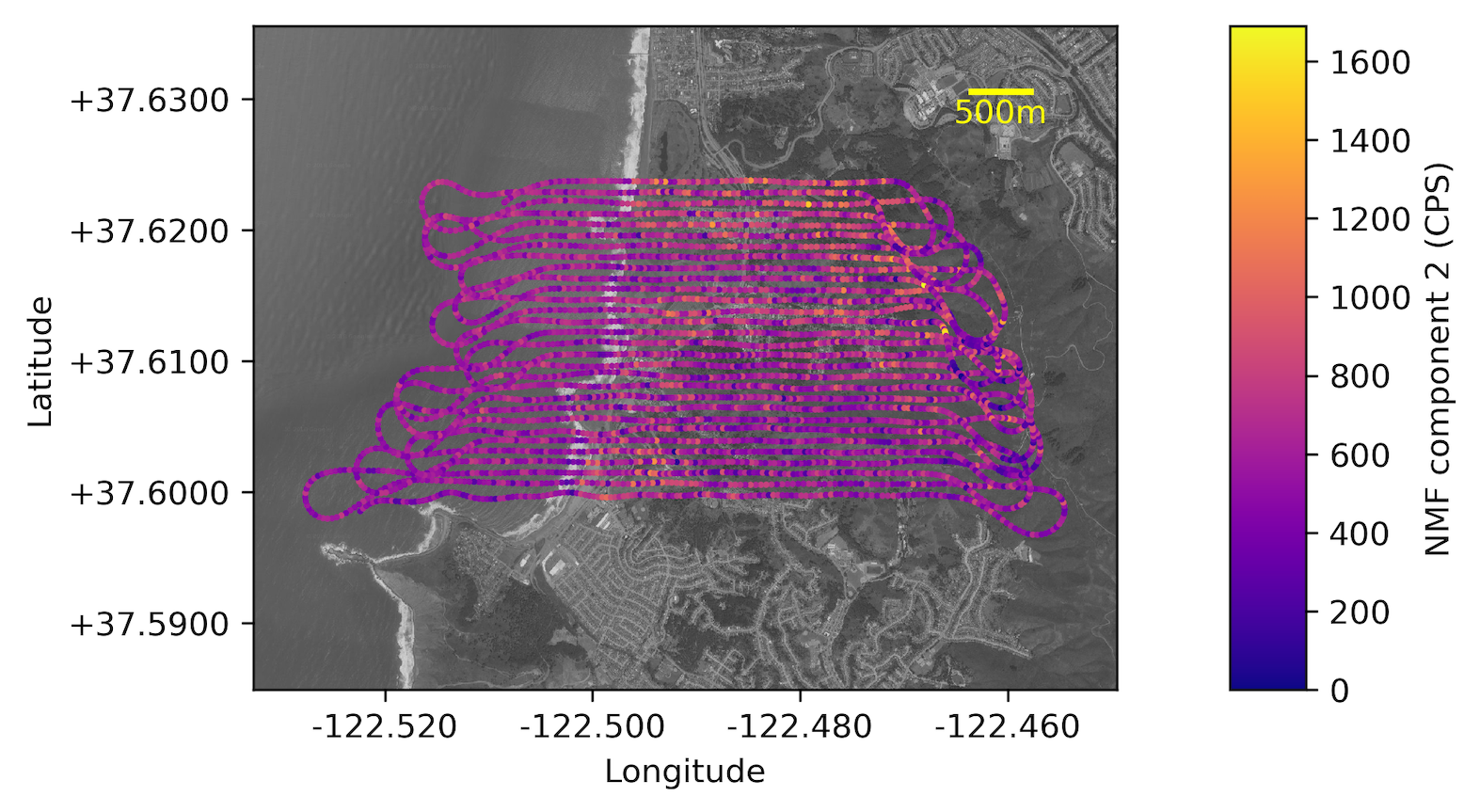}
\caption{Maps of the weights for each of the three NMF components in the Port of Oakland (left) and Pacifica (right) datasets.
For both NMF decompositions, component~0 (top) has sharper land-water boundaries than component~1 (middle), while component~2 (bottom) remains relatively constant.
(Map imagery: Google.)\label{fig:bayarea_maps}}
\end{figure*}

Maps were generated for both datasets by plotting the NMF component weights on a map without any further corrections (\Fref{fig:bayarea_maps}).
As expected from the reasoning developed in~\Fref{sec:lakemohave}, component~0 has more well defined land-water boundaries than component~1, consistent with component~0 representing nearby terrestrial emission and component~1 representing distant terrestrial emission.
Component~2 is flat for the entirety of both surveys, which is consistent with cosmic and radon emission.

\subsection{Modeling the source terms}\label{sec:bayarea_modeling}
The NMF components were modeled in a similar way as the Lake Mohave dataset (\Fref{sec:lakemohave_modeling}), with the primary difference being that the response matrices \(\mathbf{R}_{\mathrm{dir,near}}\), \(\mathbf{R}_{\mathrm{sky,near}}\), \(\mathbf{R}_{\mathrm{dir,dist}}\), and \(\mathbf{R}_{\mathrm{sky,dist}}\) were adjusted for the higher altitude using the same suite of simulations.
In an attempt to model a potential \isot{K}{40} background from the aircraft, an extra \isot{K}{40} source was fit to component~2 that consisted of \(\mathbf{R}_{\mathrm{abs}}\) applied to a \isot{K}{40} spectrum.
The best value of \(r_{\mathrm{cutoff}}\) was determined to be \(145\)~m for the Port of Oakland and \(135\)~m for Pacifica (cf. \(r_{\mathrm{cutoff}} = 85\)~m for the Lake Mohave dataset at \(100\)~ft).
The optimal power-law index was found to be~\(1.20\) for both datasets, significantly larger than the Lake Mohave value (\(0.65\)) but closer to the measurement in ref.~\cite{sandness_accurate_2009} (\(1.3\)).
It is unknown why there is such a large discrepancy between the Lake Mohave power-law index and the Bay Area power-law index, but it could point to a deficiency in the modeling of the radon, cosmics, or both.

\begin{figure*}[t!]
\centering
\includegraphics[width=0.4\textwidth]{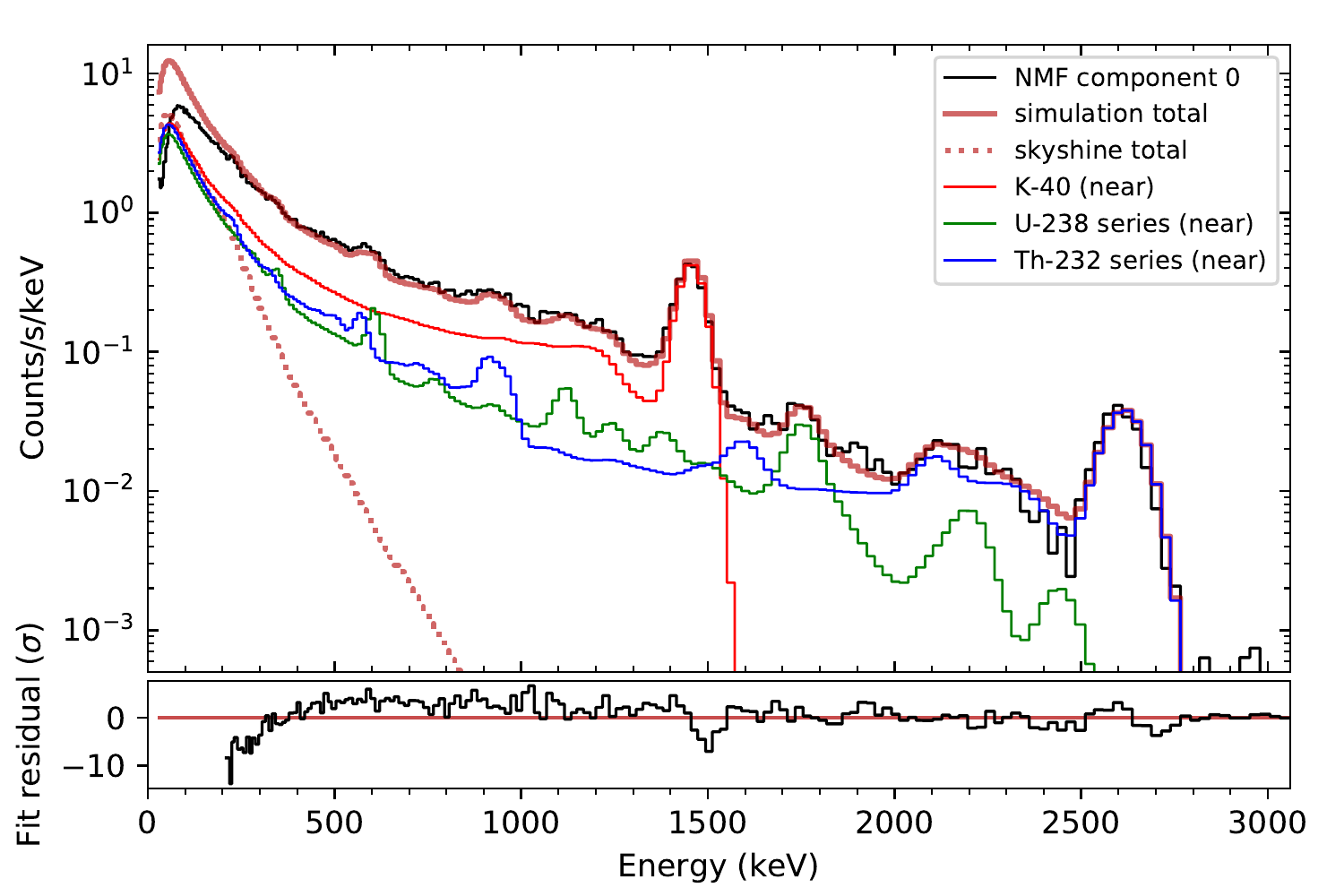}
\includegraphics[width=0.4\textwidth]{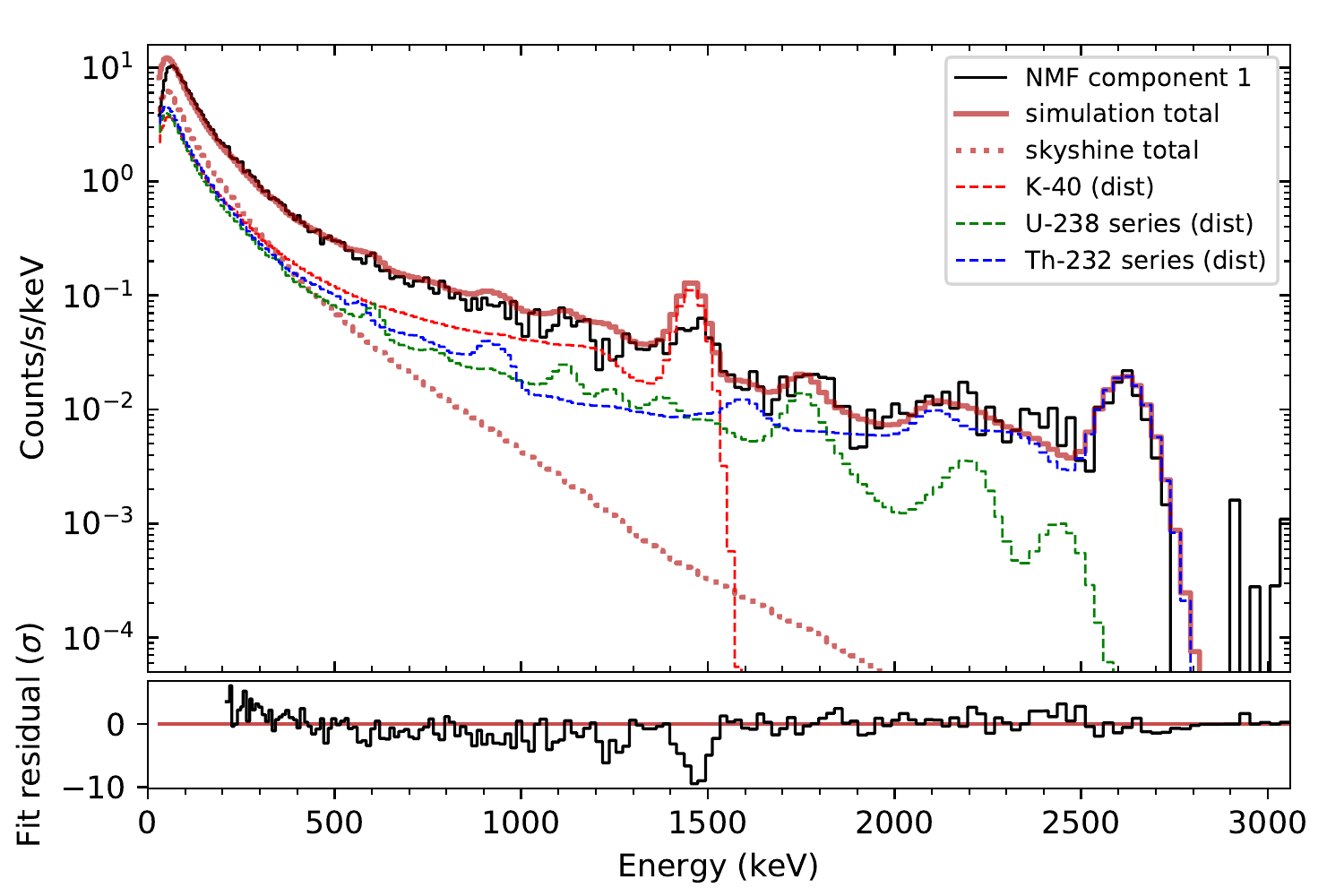}\\
\includegraphics[width=0.4\textwidth]{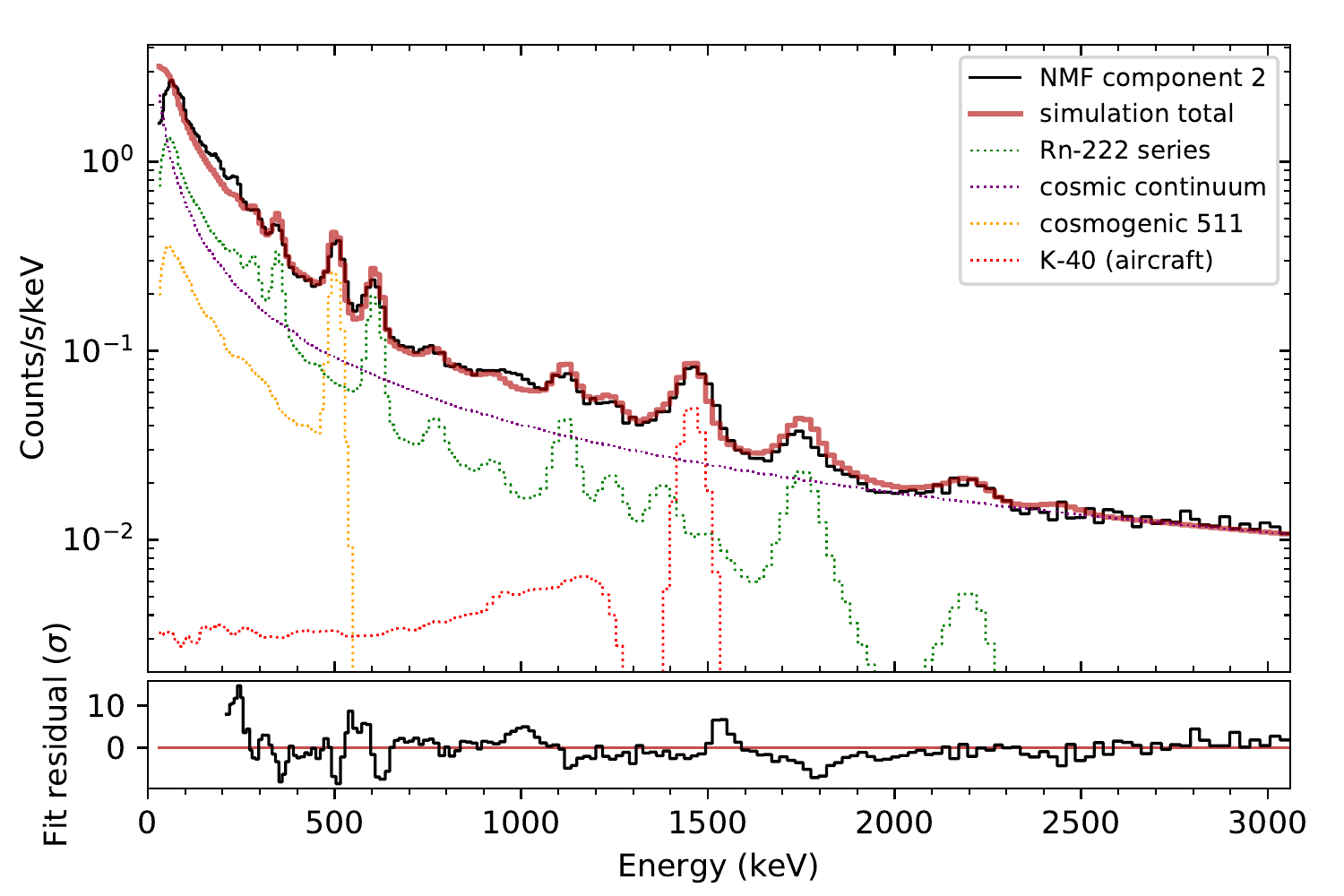}
\includegraphics[width=0.4\textwidth]{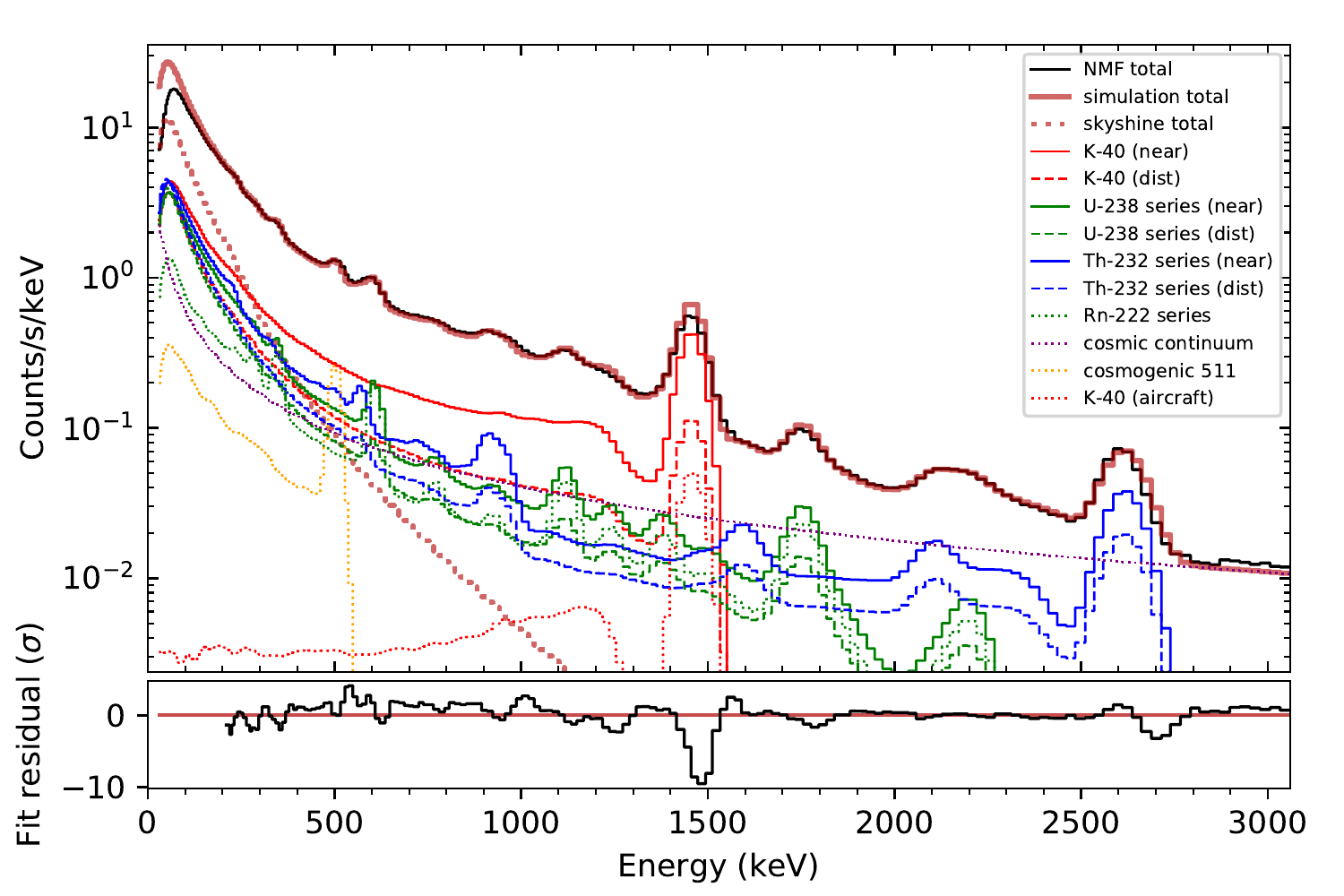}
\caption{Monte Carlo background components fit to the three NMF components (top left, top right, and bottom left) and the sum of all three components (bottom right) from the Port of Oakland dataset.
The skyshine contributions from the terrestrial sources are shown separately.\label{fig:bayarea_fits}}
\end{figure*}

The results of fits between the Monte Carlo background components and NMF components for the Port of Oakland are shown in \Fref{fig:bayarea_fits}.
These fits were constrained in the same way as the fits to the Lake Mohave components.
Once again, the modeled backgrounds are a qualitatively good fit to the NMF components, providing evidence for the identification of component~0 with nearby emission, component~1 with distant emission, and component~2 with radon and cosmic emission at this altitude.

As in the Lake Mohave fits, the model for component~0 has an excess continuum below \(100\)~keV, although these energies could easily be attenuated by a small amount of material near the detectors.
Unlike the Lake Mohave fits, the model for component~2 also has an excess below \(100\)~keV due to the larger contribution of the cosmic continuum relative to radon (possibly due to the higher altitude above ground level).
This excess could also be attenuated by material near the detectors, which is not included in this model.


\section{Discussion}\label{sec:discuss}
In this work we have presented the application of NMF to the decomposition of gamma-ray spectra from aerial surveys, with a particular focus on identifying background source terms for the resulting spectral components.
Since NMF uses the data itself to derive the primary spectral components, it has the advantage over FSA of being able to adapt to spectral shapes that have not been included in modeled components.
When compared to NASVD, NMF is able to reconstruct spectra to a similar fidelity while preserving both non-negativity and maximizing Poisson likelihood, and, in at least the cases presented here, the components appear to have a plausible physical origin.

Here we have focused on aerial data that includes water, a feature that isolates the radon and cosmic background components and favors a three-component NMF decomposition.
If the data do not include time over water, it may be harder to separate radon and cosmics from other background.
For these situations, other methods such as altitude spirals, where the aircraft flies repeated patterns in the same location at increasing altitudes, may help separate cosmics and radon from other background.
Also, a validated Monte Carlo model could potentially be used to initialize an NMF decomposition in order to guide it toward physically meaningful components.

A feature of NMF decompositions in the cases presented here is that NMF can approximately separate distant terrestrial emission from other background sources using the data alone.
Evidence for this separation has also been seen in vehicle-borne gamma-ray data~\cite{bandstra_attribution_2018}.
By exploiting this separation, NMF can potentially improve the resolution of aerial survey maps by disentangling distant emission from nearby emission.

In addition, by finding background components with physical origins, NMF could be leveraged in new algorithms for anomaly detection and background estimation.
NMF has already been shown to be competitive with PCA-based methods for spectral anomaly detection, which may be due to its accurate treatment of Poisson statistics and ability to be consistent with physics~\cite{bilton_non-negative_2019}.
The different temporal variability of the components could be exploited by Kalman filters or low-pass filters to find anomalous behavior.
Even in the data presented here from the Port of Oakland (\Fref{fig:bayarea_weights}), the anomalies found by the SCRAD metric can be easily identified as brief spikes in the NMF component~1 count rate.
Since component~1 is believed to arise from distant emission sources, it should not exhibit such high frequency behavior.
Component~2 also increases during the large anomaly at index \(1964\), which is an unusual departure from its relatively constant count rate.

With the ability to decompose the low-energy region of the spectrum where skyshine and down-scatter are most significant, NMF could be used by algorithms intended to identify anomalies below \(400\)~keV\@.
This regime is especially of interest when searching for spectral anomalies due to nuclear material, since many important isotopes have gamma-ray lines below \(400\)~keV (e.g.,~\cite{martin-burtart_airborne_2012}), however that region of the spectrum is notoriously difficult to model.

One drawback with using NMF is that the training can be computationally intensive --- the \(k=3\) model took about \(100\)~s running on four \(2.7\)~GHz cores, however the matrix computations can easily be split across any number of available cores.
\Fref{tab:compute} shows the duration of the training time for different NMF models trained on the Lake Mohave dataset, including the time it takes to calculate the weights for a single spectrum after the training has been completed.
The time to calculate the NASVD decomposition is shown for comparison.
The length of training time is set largely by the arbitrary convergence criterion used here, and NMF models in general may not need to meet the same criterion to be useful.

\begin{table}
\centering
\caption{Computation times for NMF on the Lake Mohave dataset.\label{tab:compute}}
\begin{tabular}{ccrr}
    Method & Components & Training & Single spectrum \\
    \midrule
    NASVD & N/A & \(0.44\)~s & N/A \\
    NMF & \(k=1\) & \(0.70\)~s & \(2.1\)~ms \\
    NMF & \(k=2\) & \(65.81\)~s & \(10.2\)~ms \\
    NMF & \(k=3\) & \(107.13\)~s & \(11.7\)~ms \\
    NMF & \(k=4\) & \(389.74\)~s & \(14.4\)~ms \\
    NMF & \(k=5\) & \(1,347.28\)~s & \(17.6\)~ms \\
\end{tabular}
\end{table}

Even though the training can take a significant length of time, once the NMF model has been trained for a given detector and environment and \(\mathbf{V}\) is fixed, the subsequent calculation of weights for previously unseen spectra is fast.
This application highlights another distinction between FSA, NMF, and NASVD --- typically, FSA is used to decompose spectra that have not yet been measured, while NASVD is used to decompose only previously measured spectra (although see ref.~\cite{kulisek_real-time_2015} for a real-time application of NASVD).
NMF has been presented here retrospectively like NASVD, but both NMF and NASVD components can be fit to future measurements under the assumption that the components derived from the training set are also effective at reducing the dimensionality of future data.
Using a trained NMF model to decompose spectra has been shown to be an effective tool for spectral anomaly detection in a mobile detection system~\cite{bilton_non-negative_2019}.

Finally, NMF is a potentially powerful framework for any analysis involving variable gamma-ray backgrounds since it provides a consistent description for both the mathematical and physical characteristics of the measurements.
As such, NMF provides a natural framework for data fusion, such as incorporating non-radiological contextual information about the environment.
Some work on vehicle backgrounds has already shown evidence for the ability to connect NMF components to non-radiological contextual information (i.e., video images)~\cite{bandstra_attribution_2018}.
As a simple example of data fusion for aerial measurements, having knowledge of the distance to a land-water interface would provide useful information about the behavior of the distant and nearby background components.
This area has much promise for further research.


\section{Acknowledgments}
This work has been supported by the US Department of Homeland Security, Domestic Nuclear Detection Office, under IAA HSHQDC-11-X-00380\@. This support does not constitute an express or implied endorsement on the part of the Government.

This work was also performed under the auspices of the US Department of Energy by Lawrence Berkeley National Laboratory under Contract DE-AC02-05CH11231.
The project was funded by the US Department of Energy, National Nuclear Security Administration, Office of Defense Nuclear Nonproliferation Research and Development (DNN R\&D).

We would like to thank Ren Cooper and Joseph Curtis for their helpful comments and Erika Suzuki for proofreading this manuscript.


\bibliographystyle{IEEEtran}
\bibliography{aerial_nmf}

\end{document}